# A rate-and-state friction based criterion for the probability of earthquake fault jumps

Sylvain Michel[1,2,3,4], Oona Scotti[1,5], Sebastien Hok[1,5], Harsha S. Bhat[4], Navid Kheirdast[4], Pierre Romanet[6,2], Michelle Almakari[4], Jinhui Cheng[4,7]

[1] Institut de Radioprotection et de sûreté Nucléaire, 31 avenue de la Division-Leclerc, 92262, Fontenay-aux-Roses, France

[2] Université Côte d'Azur, CNRS, IRD, Observatoire de la Côte d'Azur, Géoazur, Sophia-Antipolis, France

[3] CNRS-INSU, Institut des Sciences de la Terre Paris, Sorbonne Université, ISTeP UMR 7193, 75005 Paris, France

[4] Laboratoire de Géologie, Département de Géosciences, Ecole Normale Supérieure, PSL Université, CNRS UMR 8538, 24 Rue Lhomond, 75005, Paris, France.

[5] Now at Autorité de Sûreté Nucléaire et Radioprotection, 31 avenue de la Division-Leclerc, 92262, Fontenay-aux-Roses, France

[6] Department of Earth Sciences, La Sapienza University of Rome, Rome, Italy

[7] Now at Division of Geological and Planetary Sciences, California Institute of Technology, Pasadena, USA

## Key points:

- Development of a rate-and-state based criterion to compute probabilities of ruptures jumping from one fault to another

- Application of the criterion in seismic cycle simulations accurately predicts jumps while Coulomb stress criterion is insufficient

- The maximum jump distance predicted by the criterion increases non-linearly with decreasing normal stress (i.e. depth)

## Abstract

Geometrical complexities in natural fault zones, such as steps and gaps, pose a challenge in seismic hazard studies as they can act as obstacles to seismic ruptures. In this study, we propose a criterion, which is based on the rate-and-state equation, to estimate the efficiency of an earthquake rupture to jump between two spatially disconnected faults. The proposed jump criterion is tested using a 2D quasi-dynamic numerical simulations of the seismic cycle. The criterion successfully predicts fault jumps where the simpler Coulomb stress change calculation fails to do so. The criterion includes the Coulomb stress change as a parameter but is also dependent on other important parameters among which is the absolute normal stress on the fault the rupture jumps to. Based on the criterion, the maximum jump distance increases with decreasing absolute normal stress, i.e. as the rupture process occurs closer to the Earth's surface or as pore pressure increases. The criterion implies that earthquakes can jump to arbitrary large distances at the Earth's surface if the normal stress is allowed to go to zero, underscoring the potential for large jump distances (i.e. >5 km). We further propose a probabilistic framework to estimate the likelihood of rupture jumps by accounting for uncertainties in fault geometry and earthquake source parameters. Additionally to its role into seismic hazard assessment, this criterion could complement Coulomb stress change maps with those of triggered slip-rates on receiver faults due to quasi-instantaneous stress perturbations, as well as estimates of jump probabilities accounting for parameter uncertainties.

## Plain language summary

This study evaluates the physical conditions that allow earthquake ruptures to jump across geometrical complexities of faults, like steps and gaps. These geometrical complexities can sometimes stop an earthquake from propagating along a fault. The approach taken here is based on the *rate-and-state* equation which describes the behavior of a fault from a set of frictional parameters, considering the case of two separated fault segments. The approach was tested successfully on numerical simulations of seismic cycles, providing consistently better results than the *Coulomb stress change* approach. Indeed, this new approach allows us to better take into account the effects of the normal stress on the fault that the earthquake is jumping to. It suggests that earthquakes can likely jump farther if the normal stress on the fault is lower, which happens for example as you get closer to the Earth's surface. The way normal stress is considered is thus an important factor. This study also propose a way to calculate the likelihood of an earthquake to jump across fault geometrical complexities based on uncertainties in the faults' geometry.

**Keywords**: Earthquake Rupture Fault jump, seismic cycle simulation, rate-and-state friction, steps and gaps, Coulomb stress change, physics based approach

# 1. Introduction

Evaluating the efficiency of an earthquake rupture to jump from one fault to another is fundamental to anticipate the maximum magnitude of earthquakes (Harris et al., 1991). Is there a maximum distance between two faults an earthquake cannot jump? Can we quantify the probability of jumping based on fundamental parameters controlling fault interactions?

Multiple types of geometrical obstacles opposing rupture propagation exist such as fault bends, steps, or jogs (Biasi and Wesnousky, 2021, 2016; Oglesby, 2004; Sibson, 1985). In this study, we will focus on a setting where two faults are disconnected spatially but can potentially rupture together. This regroups geometrical obstacles described as gaps and steps (Figure 1), which are defined here as the parallel and orthogonal distance between two disconnected faults, respectively (Biasi and Wesnousky, 2016).

Many earthquakes have demonstrated that seismic ruptures can jump across fault gaps or step overs. Based on surface observations, Biasi & Wesnousky (2016) and Wesnousky (2008) have documented which earthquake succeeded or failed to pass such geometrical complexities and have also quantified their numbers and sizes. For example, the 1992 $M_w$7.2 Landers, California earthquake is just one event among many that succeeded in jumping multiple steps: three of 1.5, 2 and 3 km distance in this case. Based on their observations, Biasi & Wesnousky (2016) suggested that seismic events are not able to jump steps beyond 5 km, an upper bound considered in many seismic hazard analysis (e.g. UCERF 3; Field et al., 2014 ; Scotti et al., 2019). However, some recent events were suggested to have jumped across greater distances, such as the 2016 M7.8 Kaikoura earthquake in New Zealand (jump of an apparent restraining step of 15 km; Diederichs et al., 2019; Hamling et al., 2017; Ulrich et al., 2019). While the number of events within the catalogs increase with each study (Baize et al., 2020; Rodriguez Padilla et al., 2024; Sarmiento et al., 2024), these observations are naturally limited as they are collected at the surface, specific to each site and essentially blind to the details of how and why rupture jump or do not jump (e.g. fault geometry and rupture history at depth).

Numerous studies have used fully dynamic numerical simulations to understand further the physics behind the efficiency of an earthquake to jump from one fault to another (e.g. Bai and Ampuero, 2017; Duan and Oglesby, 2006; Harris and Day, 1999, 1993; Hu et al., 2016; Kroll et al., 2023; Lozos et al., 2012; Mia et al., 2024; Oglesby, 2008; Ryan and Oglesby, 2014; Shaw and Dieterich, 2007). Most studies simulate one single event and rely on the slip weakening friction law in which the evolution of friction with slip is predefined (e.g. Bai & Ampuero, 2017; Harris & Day, 1999). Recent studies have examined the effects of bi-material on rupture jumps (Hu et al., 2025) as well as the potential role of a small intermediate fault located between the two main faults of the stepover (Lozos et al., 2012). Some studies concentrate on replicating the rupture of complex 3D fault networks due to a single known event, as for example the 2023 Kahramanmaraş M7.8 and 7.7 earthquakes (Gabriel et al., 2023), or the 1812 co-rupture of the San Andreas and San Jacinto faults (Lozos, 2016). While those studies are insightful, they rely on an initial stress distribution of faults, constrained by information from kinematic models, and fixed prior to the simulation, which then determines whether a seismic rupture will propagate to a neighboring fault or stop. The ratio between strength excess and stress drop is often used as a criterion to quantify the relative prestress level and assess the triggering potential of a fault (Andrews, 1976; Das and Aki, 1977), although the parameters involved are difficult to quantify. On the other hand, seismic cycle simulations allow us to model sequences of earthquakes of different sizes as well as the period in between events (e.g. pre-, inter-, and post-seismic periods) (Duan, 2019; Duan and Oglesby, 2005; Erickson et al., 2020; Jiang et al., 2022; Lapusta et al., 2000). They have similarly been used to study the effect of geometrical complexities on rupture propagation (e.g. Duan et al., 2019; Liu et al., 2022; Ozawa et al., 2023). A majority of those models rely on the empirical rate-and-state friction law (Dieterich, 1979, 1972) which is expressed as:

$$\tau = \sigma \left[ f_* + a\, ln\left(\frac{V}{V_*}\right) + b\, ln\left(\frac{V_*\, \theta}{L}\right) \right], \tag{1}$$

where $\tau$ and $\sigma$ correspond to the shear and normal stress, respectively, $V$ is the slip rate and $f_*$ is the coefficient of friction at the reference slip rate $V_*$. $L$ is a characteristic slip distance. $a$ and $b$ are frictional parameters where $a$ describes the effect of shear stress in response to an abrupt change in slip rate, called the direct effect, and $b$ governs the evolution of the state variable, $\theta$. The rate-and-state friction law has to be complemented by an evolution law for the state variable $\theta$ (e.g. the aging or the slip law; (Beeler et al., 1994; Roy and Marone, 1996; Ruina, 1983). Regions where $a - b < 0$ are said to be Velocity Weakening (VW) and can potentially host earthquakes, while regions where $a - b > 0$ are said to be Velocity Strengthening (VS) and tend to creep steadily. Simulations using the rate-and-state friction law enable us to generate sequences of earthquakes and stress distributions on the fault that are controlled by the evolution and history of slip. Nevertheless, modeling multiple faults is computationally demanding and approximations are often applied to make the simulations faster. As an example, contrarily to fully dynamic simulations, quasi-dynamic ones approximate the inertial effect of seismic waves (e.g. Luo et al., 2017; Rice, 1993; Romanet et al., 2018). As another example, the quasi-static RSQSim algorithm (Dieterich and Richards-Dinger, 2010; Richards-Dinger and Dieterich, 2012) simplifies the rate-and-state friction law behavior into three slip regimes (i.e. locked, nucleating or dynamically slipping) and has been used to study sequences of earthquakes on complex fault networks, including rupture jumps (e.g. Shaw et al., 2022).

While the use of simulations using the rate-and-state friction law begins to be widely used to study fault network interactions, and even incorporated into seismic hazard analysis (e.g. Chartier et al., 2021; Shaw et al., 2018), the study of the fundamental parameters controlling fault interactions has not been studied in detail within this framework, in particular concerning the efficiency of rupture jumps. Moreover, for seismic hazard, it is important to explore the uncertainty of each of the rate and state parameters and of the fault properties as it allows to estimate in a probabilistic approach the efficiency of an earthquake to pass an obstacle. For instance, some studies have focused on the efficiency of ruptures to pass frictional obstacles or geometrical fault complexities (Huang et al., 2025; Kaneko et al., 2010; Molina-Ormazabal et al., 2023; Ozawa et al., 2023), but none proposed a rupture jump criterion for gaps and steps using the rate-and-state framework exploring the full space of parameters. This modeling framework can then be used to evaluate a fault's seismogenic potential (Michel et al., 2021), taking into account the effect of the obstacles, and be included into seismic hazard analysis (Biasi and Wesnousky, 2021).

Our study aims to better characterize the efficiency of an earthquake to jump from one fault to another based on the rate-and-state friction law, and identify the parameters that control fault interaction. To do so, we first build a rupture jump efficiency criterion which roots from the rate-and-state equation (Eq. [1]), and test it against numerical earthquake sequences generated along two faults using the 2D quasi-dynamic seismic cycle algorithm VEGA developed by Romanet et al. (2018). We then explore the implications of the criterion in terms of maximum jump distance and propose an approach to compute probabilities of rupture jumps. Finally, we discuss about the limits of the criterion before concluding.

## 2. Rupture Jump Efficiency Criterion

To build a rupture jump efficiency criterion, we use the rate-and-state formulation (i.e. Eq. [1]) for the slip rate parameter $V$:

$$V = exp\left[\frac{1}{a}\left[\frac{\tau}{\sigma} - f_* - \; b\; ln\left(\frac{V_*\,\theta}{L}\right)\right] + ln(V_*)\right]. \qquad (2)$$

The ratio between the slip rate on a fault prior and after the interaction of an earthquake that occurred on a neighboring fault, denoted $V_0$ and $V_i$, respectively, is thus expressed as:

$$\frac{V_i}{V_0} = exp\left[\frac{1}{a}\left[\frac{\tau_i}{\sigma_i} - \; b\; ln\left(\frac{V_*\theta_i}{L}\right)\right] - \frac{1}{a}\left[\frac{\tau_0}{\sigma_0} - \; b\; ln\left(\frac{V_*\theta_0}{L}\right)\right]\right]. \qquad (3)$$

If we assume that the earthquake interaction is instantaneous, then the state variable $\theta_0 = \theta_i$ and:

$$\frac{V_i}{V_0} = exp\left[\frac{1}{a}\left(\frac{\tau_i}{\sigma_i} - \frac{\tau_0}{\sigma_0}\right)\right]. \qquad (4)$$

This expression can be reordered as follows:

$$V_i \; = \; V_0\; exp\left[\; \frac{1}{a\,\sigma_i}\; \left((\tau_i - \tau_0) - \frac{\tau_0}{\sigma_0}(\sigma_i - \sigma_0)\right)\; \right]. \qquad (5)$$

$V_i$ is thus the absolute value of the slip rate on the fault after a stress interaction, and will be used as the rupture jump efficiency criterion and tested in the next section on seismic cycle simulations. We rename it $V_{RSSP}$ for *Rate-and-State Stress Perturbation* (RSSP). $(\tau_i - \tau_0)$ and $(\sigma_i - \sigma_0)$ correspond to the shear and normal stress changes on the fault due to the earthquake on a neighboring fault, while $\frac{\tau_0}{\sigma_0}$ correspond to the effective coefficient of friction before the earthquake. Equation [5] can then be expressed as:

$$V_{RSSP} = V_0 \, exp\left[ \frac{1}{a\,(\sigma_0 + \Delta\sigma)} \left( \Delta\tau - \frac{\tau_0}{\sigma_0} \Delta\sigma \right) \right] \approx V_0 \, exp\left[ \frac{\Delta C}{a\,(\sigma_0 + \Delta\sigma)} \right], \quad (6)$$

where $\Delta C$ is the Coulomb stress change (Reasenberg and Simpson, 1992). Notice that a slightly different formulation is mentioned in the study from Kroll et al. (2023) (Text S1).

The definition of an earthquake in rate-and-state cycle simulations is ambiguous since in this framework the faults never stop slipping and slip rates can vary in orders of magnitude. Here we define it as any portion of a fault with a slip rate above $10^{-3}$ m/s (Romanet et al., 2018). Thus, if an earthquake interaction on a second fault increases $V_{RSSP}$ above $10^{-3}$ m/s, this earthquake has, by our definition, jumped.

$V_{RSSP}$ implies that the instantaneous reaction of the fault an earthquake tries to jump to does not solely depend on the Coulomb stress change (i.e. $\Delta C$). All the parameters of $V_{RSSP}$ are related to the fault the rupture tries to jump to, however, the information about the size of the rupture trying to jump and the geometry of the faults are included in $\Delta C$ and $\Delta\sigma$. The other parameters, $V_0$, $a$, and $\sigma_0$, modulate the spatial pattern implied by $\Delta C$ and $\Delta\sigma$. Finally, note that the criterion is only dependent on the rate-and-state parameter $a$, not on $b$. It means that, given suitable stress perturbation, an earthquake can mechanically increase the slip rate on a velocity-strengthening zone up to seismic velocities. However, the earthquake generated on the fault would likely die quickly and not propagate further. More generally the criteria does not provide any information on whether rupture initiation on a nearby fault will propagate further and become self-sustained.

## 3. Seismic Cycle Simulations

We test the predictivity of Eq. [6] based on 2D quasi-dynamic seismic cycle simulations generated from the algorithm VEGA (Romanet et al., 2018). This algorithm, based on the rate-and-state friction law, allows us to model sequences of earthquakes on a network of faults. The aging law is used here to describe the evolution of the state variable (Ruina, 1983). The description of the model geometric setting is shown in Figure 1. To represent the simplest geometry of a fault step, we model two linear parallel faults of same length, respectively Fault 1 and 2, which are separated in space. The parallel and perpendicular distance between the closest tips of the two faults, respectively $D$ and $H$, are taken relative to the direction of the first fault. Each fault is divided into a VW and VS area of identical size. The faults are loaded using a back-slip rate approach (Heimisson, 2020; Savage, 1983; Shaw et al., 2022; Tullis et al., 2012), which enables us to load each fault with different slip rates, contrary to a regional stress loading. Faults 1 and 2 are loaded at 30 and 7 mm/yr, respectively (Text S2). Fault 1 generates earthquakes at a greater frequency than Fault 2 and tests the second fault at different times of its seismic cycle, i.e. with different levels of stress distribution. Faults 1 and 2 are defined as the source and receiver faults, respectively. Without this contrast of loading, the two faults tend to synchronize and rupture together, which makes it difficult to explore a variety of scenarios. The parameters $a$ and $b$ of the friction law for the VW area are chosen equal to 0.01 and 0.018, respectively, standard values used in the literature (e.g. Michel et al., 2017). For simplicity, the parameters $a$, $b$, and $L$ are the same for both faults, with these parameters remaining constant within the respective VS and VW regions.

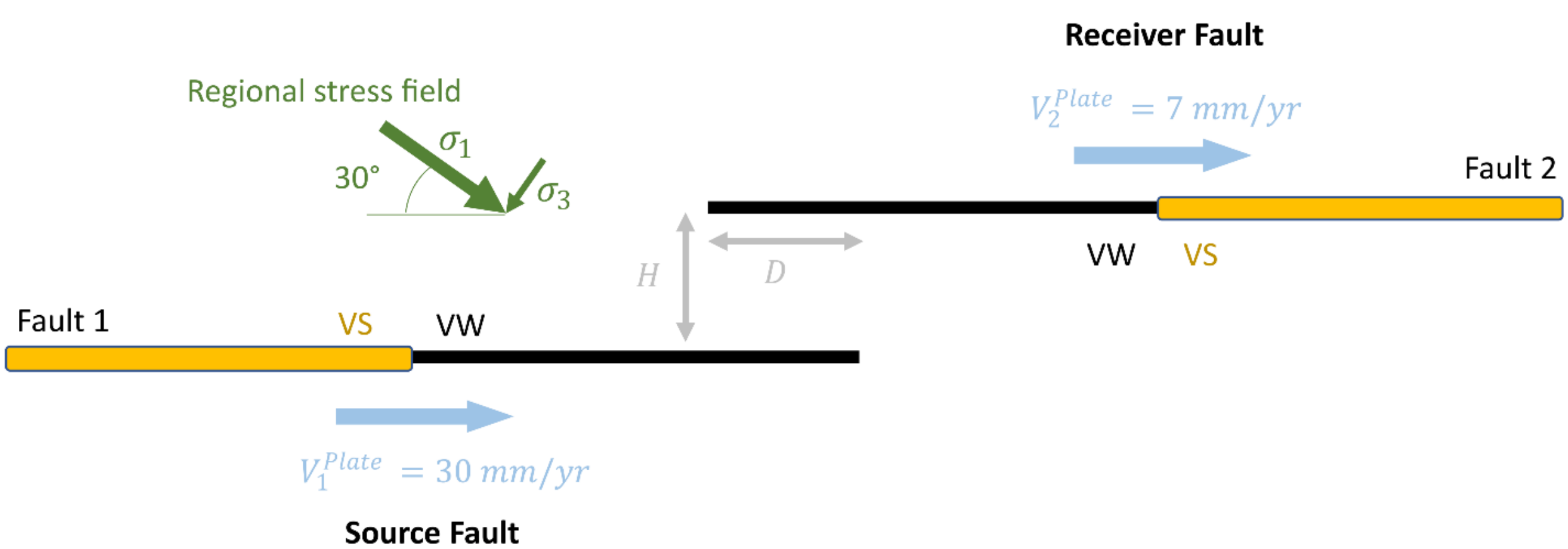

**Figure 1**: Simulations' general setting representing two faults separated in space. Different loading rates are applied to Fault 1 and 2, $V_1^{Plate}$ and $V_2^{Plate}$, respectively. The more frequent ruptures occurring on Fault 1, called the source fault, will sometime jump onto Fault 2, called the receiver fault. $D$ and $H$ correspond to the parallel (i.e. overlap) and perpendicular (i.e. step) distance between the two faults, respectively. VW and VS stand for Velocity Weakening and Strengthening, respectively. The regional stress field is set so that the simulations are in a strike-slip regime and with a maximum principal stress, $\sigma_1$, optimally oriented (i.e. ~30° from Fault 1).

Three different scenarios are considered in this study, for which the parameters are indicated in Table 1. Two scenarios have aligned faults and thus don't have any normal stress interaction but have different values of normal stress: 80 and 40 MPa, respectively. The normal stress is controlled by the regional static stress field which is optimally set at 30° from the faults (Anderson, 1905). The third scenario has the same parametrization as the first scenario except that the faults overlap a third of their VW areas and that normal stress interactions occur. For each scenario we test different lengths of the fault's VW portion (e.g. 10, 30 and 60 km), as well as different values of $D$ (from -1.5 to 20 km) and $H$ when there is an overlap (-2 to 0.5 km) (see Table 1 and Text S3). The characteristic distance $L$ of rate-and-state friction law is set to 0.0155 m which prescribes a nucleation length of 1 and 2 km for normal stresses of 80 and 40 MPa, respectively. The size of the sub-patches of the faults are determined so that they are roughly ten times smaller than the cohesive zone (Day et al., 2005; Rubin, 2002). The simulations are thus well resolved (Lapusta and Liu, 2009).

**Table 1:** List of Physical parameters used in the models for each scenario in Section 3. The fault's full length is always twice the size of the VW region.

| | **Scenario 1** | **Scenario 2** | **Scenario 3** |
|---|---|---|---|
| Main information | Aligned faults<br>$\sigma = 80$ MPa | Aligned faults<br>$\sigma = 40$ MPa | Overlapping faults<br>$\sigma = 80$ MPa |
| $a$ | 0.01 | | |
| $b$ | 0.018 | | |
| $L$ | 0.0155 m | | |
| $f_*$ | 0.6 | | |
| $V_*$ | $10^{-9}$ m/s | | |
| $\sigma$ | 80 MPa | 40 MPa | 80 MPa |
| Length of VW region | 10, 30 and 60 km | 30 and 60 km | 30 and 60 km |
| $D$ | -0.5, -1.0 and -1.5 km | -0.25 and -0.5 km | 10 and 20 km |
| $H$ | 0 km | 0 km | 0.5, -0.25, -0.5, -1 and -2 km |
| Nucleation length | ~1 km | ~2 km | ~1 km |
| Cohesive zone | ~360 m | ~730 m | ~360 m |
| Sub-patches size | 36 m | 61 m | 36 m |

# 4. RESULTS

## 4.1. Example of simulation

We show in Figure 2 an example of a simulation from the third scenario, with an overlap between the two faults and a perpendicular distance of 150 m (below the cohesive zone size of ~310 m), to illustrate the complexity of behavior within one simulation. Figure 2.a represents the maximal slip rate, $V_{max}$, through time occurring on each fault. The timing and spatial extent of earthquakes are defined by the detection threshold of $10^{-3}$ m/s (Romanet et al., 2018). In this simulation, spanning ~1500 yrs (3 x $10^5$ time steps), 47 events have occurred on both faults, 26 on Fault 1, among which 10 have jumped from Fault 1 to Fault 2 (numbered from 1 to 10 in Figure 2.b). Figure 2.b shows the timing and size of each event on Fault 1 and indicates which have jumped (dark blue bars). Magnitudes are estimated assuming square areas for the 2D ruptures (i.e. rupture length is squared). Figure 2.c shows the spatio-temporal distribution of slip rate on both faults. Note that the time is expressed in time steps. In the simulations, time steps decrease when slip rate increases, hence the greater number of time steps during earthquakes compared to the inter-seismic period, that help visualizing the propagation of seismic events. The following comments illustrate the complexity of the fault behavior that is present within each simulation. Events nucleate on both the transition between VW and VS zone, and at the end of the fault on the VW section (e.g. jumping event 6 and 10), but also near the location of the extent of the overlap (e.g. jumping event 9). Full and partial ruptures of the VW are observed. An increase of slip rate propagating along the receiver fault during the propagation of an earthquake occurring on the source fault is also observed (e.g. blue line within the overlap zone for jumping event 6) but is only visualized for cases when a rupture is very close to the receiver fault (less than roughly the size of the cohesive zone). We note the special case of jumping event 8 which does so right after a full rupture of Fault 2, and re-ruptures a portion of Fault 2 during its post-seismic period. Finally, in Figure 2.d, we show the slip distribution of all events which are roughly parabolic or truncated parabolas. The 1$^{st}$ events on both faults of the simulations have generally larger slip amplitude due to the initial stress distribution imposed and are thereafter not taken into account in our analysis. Neither are the last events which might have been cut in time at the end of our simulations.

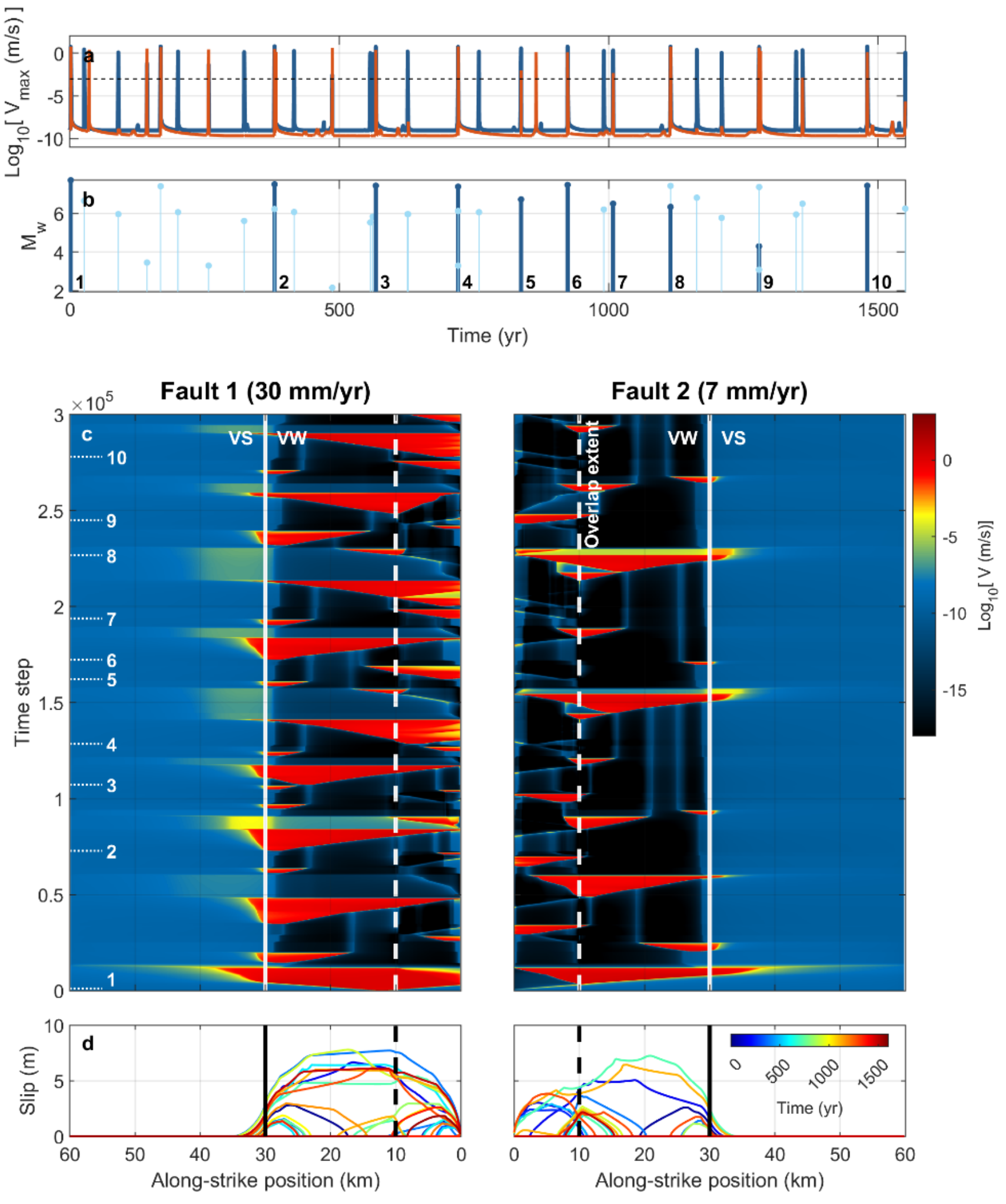

**Figure 2**: Example of simulation from two overlapping faults, i.e. within Scenario 3 (see Section 3 and Table 1). Fault 1, the source fault is loaded at 30 mm/yr while Fault 2, the receiver fault, is loaded at 7 mm/yr. (a) Maximum slip rate on Fault 1 (blue) and 2 (orange) through time. (b) Magnitude of events on Fault 1 through time. Events that jumped from Fault 1 to 2 are indicated by dark blue bars and numbered from 1 to 10, while the ones that failed to jump are in light blue. (c) Slip rate of Fault 1 and 2 through time. The time is indicated in time steps. In the simulations, time step size decreases when slip rate increases, which helps visualizing seismic events that last a few seconds in a sea of inter-seismic loading. The start of each jumping event of Fault 1 is indicated on the left side of the panel. VW and VS stand for Velocity Weakening and Strengthening, respectively. (d) Slip distribution of each event. The color indicates its timing. The first event slip distribution is not shown as it is highly dependent on the initial stress imposed.

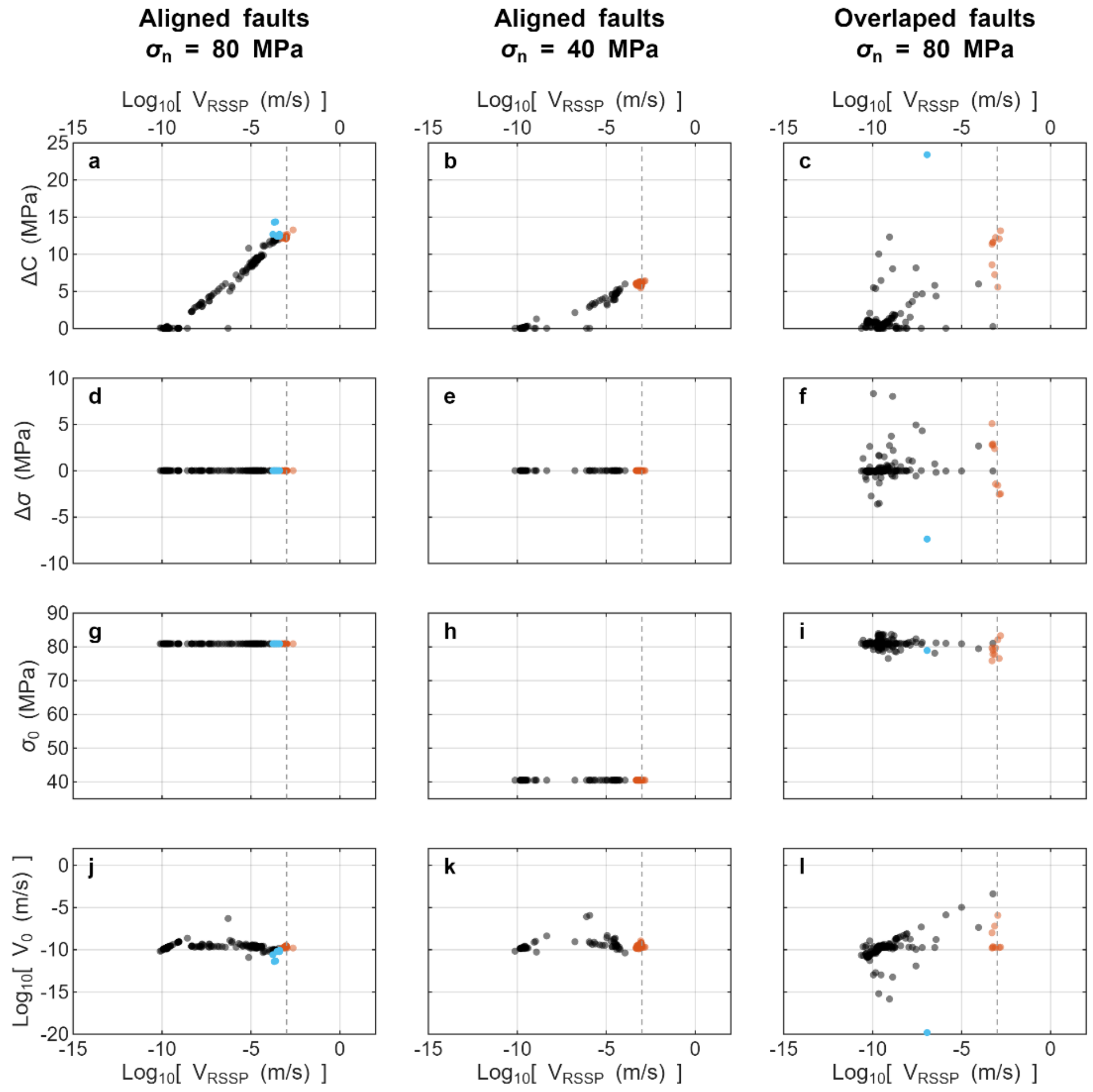

**Figure 3**: Results from the simulations for all scenarios (Section 3). For all panels, the orange and black dots indicate events that succeeded and failed to jump, respectively. Blue dots correspond to events with high Coulomb stress change, $\Delta C$, but that did not jump. (a), (b) and (c) show the $\Delta C$ on Fault 2 due to events occurring on Fault 1 at the location of maximum $V_{RSSP}$. (d), (e) and (f) show the normal stress change, $\Delta\sigma$, on Fault 2 due to events occurring on Fault 1 at the location of maximum $V_{RSSP}$. Details on how $\Delta C$ and $\Delta\sigma$ are retrieved are in Section 3. (g), (h) and (i) show the effective normal stress, $\sigma_0$, on Fault 2 just before the start of events on Fault 1 at the location of maximum $V_{RSSP}$. (j), (k) and (l) show the slip rate, $V_0$, on Fault 2 just before the start of events on Fault 1 at the location of maximum $V_{RSSP}$.

## 4.2. $V_{RSSP}$ and $\Delta C$ predictiveness in the simulations

To test the predictivity of eq. [6], we calculate the distribution of $V_{RSSP}$ along Fault 2 due to the stress impact of each earthquake $i$ on Fault 1. We thus focus only on the events generated by Fault 1 that are trying to jump on Fault 2. The parameter $a$ is fixed in our simulations. The distributions of the initial slip rate, $V_0$, normal stress, $\sigma_0$, and shear stress, $\tau_0$, on Fault 2 are sampled at the start of earthquake $i$ on Fault 1. The distributions of normal and shear stress change, $\Delta\sigma$ and $\Delta\tau$, respectively, are calculated as the difference between the distributions of the normal, $\sigma_t$, and shear stress, $\tau_t$, at the start of seismic velocities on Fault 2 if the earthquake $i$ has jumped, and $\sigma_0$ and $\tau_0$, respectively. If earthquake $i$ didn't jump, $\sigma_t$ and $\tau_t$ are sampled at the end of earthquake $i$. The stress change on Fault 2 in the simulations is expected to result from a combination of two contributions: (1) a static stress change due to slip on Fault 1 (as if Fault 2 didn't exist), and (2) a stress redistribution of Fault 2 in response to the static stress change between the sampling times (i.e. between $t_0$ and $t_t$). In the simulations, the static stress change is dominant, while the stress redistribution term is negligible, as illustrated in Figure S1. For each earthquake, we select the maximum $V_{RSSP}$ from its spatial distribution along Fault 2 and sample the other parameters at the location where this maximum $V_{RSSP}$ occurs.

The results of predictivity of Eq. [6] for the three scenarios are shown in Figure 3. We see that jumping events in the simulations (red dots in Figure 3) have a $V_{RSSP}$ close to $10^{-3}$ m/s (all within 0.4 $10^{-3}$ and 2.3 $10^{-3}$ m/s), while events that did not jump (black and blue dots in Figure 3) have all a $V_{RSSP}$ below our detection threshold. Those results confirm the predictability of eq. [6] within the model framework.

For scenario 1 and 2, the faults are aligned and there is thus no normal stress interaction (Figure 3.d, e, g and h), only shear stress interactions. We see that if we only refer to the Coulomb stress change, $\Delta C$, for predicting the jumps of earthquakes, it would be insufficient (Figure 3.a and b). Earthquakes tend to jump at $\Delta C$ $= \sim$12-13 MPa for $\sigma_0 = 80$ MPa and at $\sim$6 MPa for $\sigma_0 = 40$ MPa, half less, as expected from Eq. [6]. Overlapping faults, as in scenario 3, allow for normal stress interaction (Figure 3.f and i) which further degrade any predictability based on $\Delta C$ alone (Figure 3.c).

For aligned faults (scenario 1 and 2), the location of maximum $V_{RSSP}$ is always at the tip of Fault 2 closest to Fault 1. This zone tends to creep at loading rate (i.e. $log_{10}$(7 mm/yr) = -9.7) due to high concentration of stress (e.g. Cattania, 2019), making it easier to jump compared to locked portions of the fault (i.e. $log_{10}(V) =$ -15) as expected from eq. [6]. After an earthquake ruptures the tip of Fault 2, slip rates at this location drop to locked values ($\sim log_{10}(V) =$ -15/-20 in our simulations) and then needs a period of time to come back to loading slip rate values (see example in Figure S2). During this initial 'healing' period, it is thus more difficult for an earthquake to jump. For aligned faults, it is as hard for a rupture to jump just after a small event on the receiver fault that resets slip rates to locked values, as after a large earthquake that rupture the full VW area. For any scenarios, locations of faults that tend to creep at slip rates close to loading rates, other than VS areas, are the borders of the VW regions, the tips of the faults, and the location where residual stress has been left by previous earthquakes. Any of those locations makes it easier for an earthquake to jump. This is illustrated by events that have Coulomb stresses similar or higher than the ones needed to jump (blue dots in Figure 3.a, b and c; Text S4), but failed to do so because of low $V_0$ (Figure 3.j, k and l).

## 5. Maximum rupture jump distance predicted by $V_{RSSP}$ criterion

We saw in Section 4 that eq. [6] predicts well in the simulations whether an event jumps or not on a second fault. In this section we will focus on the implications in terms of jump distance based solely on the exploration of the parameters of eq. [6], and provide some sensitivity tests. For simplicity, and as an example for a base scenario, we still assume two parallel linear faults. For the rest of this section we assume also that the slip distribution of events are uniform along strike. With a uniform slip distribution and assuming that the second tip of the source fault is too far away to have an impact on the receiver fault (although this last assumption actually depends on the fault's length; Figure S3), the Coulomb and normal stress fields have a similar pattern for any given slip value and their amplitudes are proportional to the slip. With these assumptions, the stress fields depend only on the slip value (and the angle between the two faults; see Figure S4 and S5) and not on the length of the source fault and Eq. [6] can then be simplified:

$$V_{RSSP} \approx V_0 \, exp\left[ \frac{\overline{\Delta C}\, S}{a\,(\sigma_0+\overline{\Delta\sigma}\, S)} \right], \tag{7}$$

where $\overline{\Delta C}$ and $\overline{\Delta\sigma}$ are the Coulomb and normal stress change normalized by the slip, and $S$ is the slip. Those assumptions can easily be changed if wanted. We show in the supplement an example using an elliptical slip distribution (Figure S6).

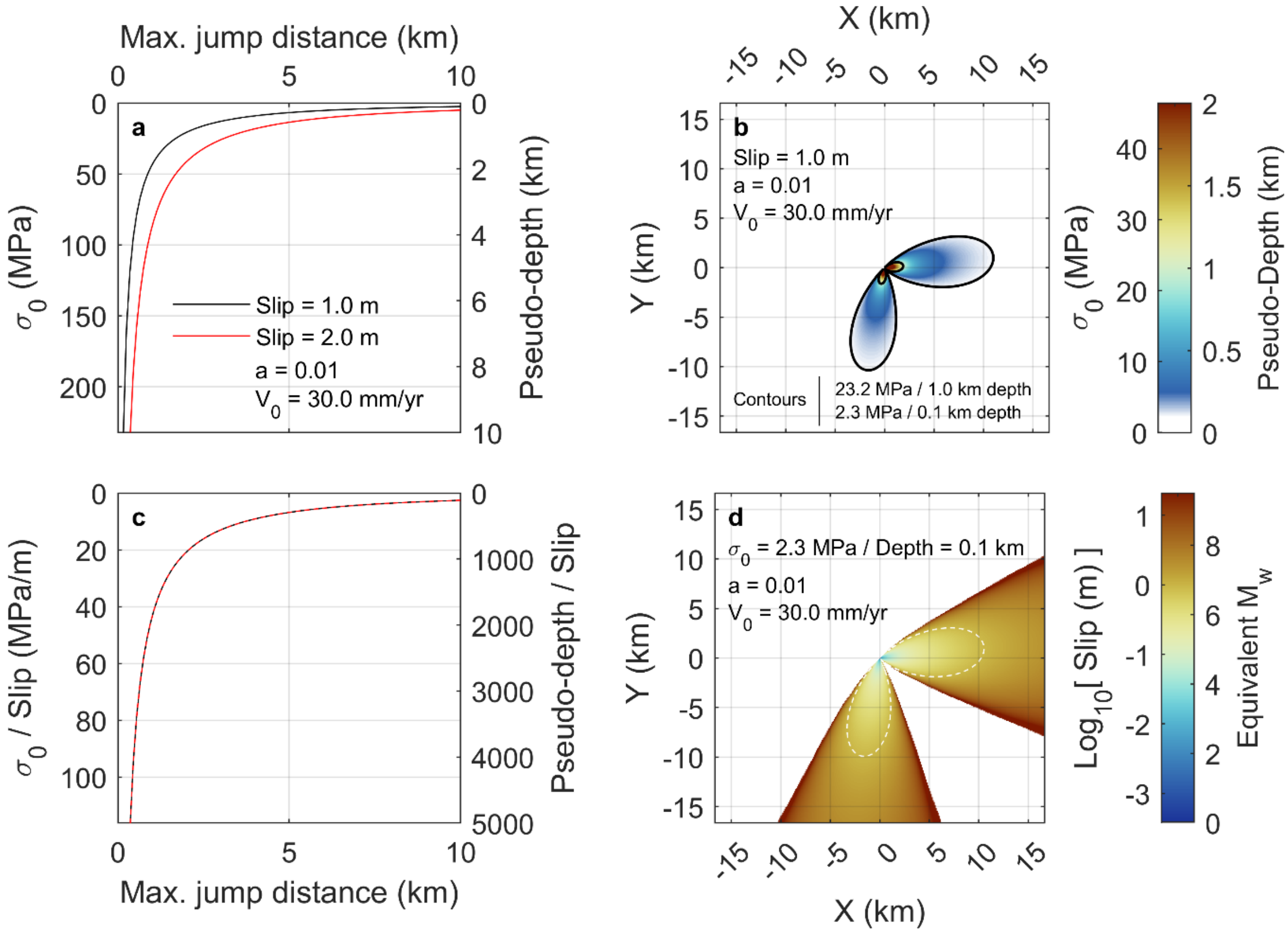

**Figure 4**: Impact on the maximum jump distance of the parameters in Eq. [6] assuming uniform slip distributions along the fault and no contamination of the stress impact from the second tip of the source fault, the one furthest away from a potential 2nd fault. All tests were realized using the same fixed values of $a$ and $V_0$. (a) profile of $\sigma_0$ along the direction of Fault 1 as a function of the maximum jump distance, for a rupture event with 1.0 and 2.0 m uniform slip. Decreasing linearly $\sigma_0$ increases exponentially the maximum jump distance, up to infinity as $\sigma_0$ approaches 0. $\sigma_0$ can be interpreted as a pseudo depth. We assume a gradient of 23 MPa/km. (b) Map of maximum jump distance for an associated $\sigma_0$ for an event of 1.0 uniform slip ($\sim M_w 6.8$). The contours indicate the position of the maximum jump distance for $\sigma_0 =$ 23 and 2.3 MPa, corresponding to a pseudo depth of 1 and 0.1 km, respectively. The profile in panel (a) is taken from this map along the coordinate Y=0. (c) Same profiles as in panel (a) but normalized by their respective slip amplitude. Both curves collapse. (d) Map of maximum jump distance for an associated slip

amplitude fixing $\sigma_0$ to 2.3 MPa (pseudo depth of ~0.1 km). The dashed white contour indicates the position of the maximum jump distance for a uniform slip of 1 m ($\sim M_w 6.8$).

We first aim to estimate the maximum jump distance as a function of $\sigma_0$, which in turn can be transcribed in terms of pseudo depth using one's favorite model. The conversion is based here on a fault in strike-slip regime and hydrostatic conditions, with a gradient of normal stress with depth equal to 23.2 MPa/km (see Text S5 for details). We effectively determine a map of $\sigma_0$ needed for an earthquake of a given slip to jump. To do so, we isolate $\sigma_0$ in eq. [6] and assume $V_{RSSP} = 10^{-3}$ m/s, $V_0 = 30$ mm/yr (i.e assuming there is always on the receiving fault an area slipping at loading rate; see last paragraph of Section 4.2), $a = 0.01$ and test a constant slip amplitude of 1.0 m, equivalent roughly to a $M_w 6.8$ (using the slip-length and length-magnitude scaling laws from Leonard (2010)). The map of $\sigma_0$ related to the maximum jump distance is represented in Figure 4.b. It shows that for relatively strong $\sigma_0$, it is relatively difficult to jump (e.g. 1.25 km maximum jump distance for $\sigma_0 = 23.2$ MPa / ~1 km pseudo depth; see contour line). Inversely, as $\sigma_0$ decreases linearly, it is exponentially easier to jump a larger distance (e.g. ~11 km maximum jump distance for $\sigma_0 = 2.3$ MPa / ~0.1 km pseudo depth; see contour line). This is better seen on Figure 4.a, in which the black curve represents a cut section of the $\sigma_0$ map for $\mathrm{Y} = 0$ km. As $\sigma_0$ approaches zero linearly, the maximum jump distance goes to infinity. If interpreted in terms of depth, maximum jump distance approaches infinity when approaching the surface. This has strong implications as it suggests that most earthquake ruptures that jumps from one fault to another should do so close to the surface. Note that the map in Figure 4.b and the profile in Figure 4.a can be easily modified to retrieve similar results for different values of slip. This is possible based on the assumptions of uniform slip distribution and negligible impact of the second tip of the source fault. Figure 4.c is an illustration of the normalization of the maximum distance of jump as a function of $\sigma_0$ based on the slip value. We want to emphasize that $\sigma_0$ represents the effective normal stress on the fault and that it is also dependent on pore fluid pressure (Huang et al., 2025). Any areas with high fluid pressure which induces low effective normal stress will also facilitate rupture jumps (Huang et al., 2025).

A similar exercise can be applied to estimate the maximum jump distance as a function of slip, which in turn can be transcribed in terms of magnitude (e.g. using the slip-length and length-magnitude scaling laws from Leonard, 2010)). If we fix $V_{RSSP} = 10^{-3}$ m/s, $V_0 = 30$ mm/yr, $a = 0.01$ and consider a value of $\sigma_0 = 2.3$ MPa (~0.1 km pseudo depth), a map of slip related to the maximum jump distance can be retrieved (Figure 4.d). As expected, a slip of ~1.0 m (~$M_w$6.8) allows for a jump of ~11 km as seen in Figure 4.b.

# 6. Computation of jump probabilities

## 6.1. Example for a fixed angle between two faults

In this section, we show how the equation for $V_{RSSP}$ can be used to compute probabilities of an earthquake to jump between two faults. For simplicity, we again assume two parallel linear faults, uniform slip distribution for seismic events and negligible impact of the stress field due to the second tip of the source fault. In this section, the angle between the faults is fixed while in the next Section we will explore the uncertainty on the angle.

To provide an example, we test a setting where the 2nd fault is at $H = 400$ m from the first fault (i.e. restraining step; Figure S7), with an overlap of $D = 3$ km, and explore uncertainties of the parameters of eq. [6]. We assume that the source fault is 40 km long (~$M_w$6.8), which, using the length-slip scaling law from Leonard (2010) and related uncertainty, produce a normal distribution of slip in $log_{10}$ scale: $N(log_{10}(1.0), 0.4)\ log_{10}(m)$ (Figure 5.e). Note that only the slip is needed to compute $V_{RSSP}$ and that the length-slip scaling law is solely used here to provide uncertainties on the slip. Those uncertainties are quite large since the 5 and 95 % percentiles of the distribution are associated with slip of 0.2 and 4.5 m. For $\sigma_0$, we assume a normal distribution of $N(23.2{,}6.0)\ MPa$, corresponding to the conditions at 1 km pseudo-depth according to the normal stress with depth gradient used in this study (Figure 5.f; see Text S5 for details). The distribution is truncated at 11.3 and 35.2 MPa at its lower and upper bounds, respectively, to keep our example in a strike-slip regime. $V_0$ is assumed Gaussian with a mean value of 30 mm/yr and an uncertainty equivalent to 10% of its mean: $N(30\ {,}3)\ mm/yr$ (Figure 5.g). This choice for $V_0$ assumes that there is always a point on the receiver fault that is at least slipping at loading rate and available for the jump (see last paragraph of Section 4.2). $a$ is also assumed Gaussian, $N(0.010{,}0.005)$, but is truncated at 0 to avoid any negative values (Figure 5.h).

These distributions are sampled 30 000 times allowing to compute 30 000 maps of $V_{RSSP}$. For each map, any location with $V_{RSSP} > 10^{-3} m/s$ is assumed to be a location were the earthquake rupture tested will jump. The probability of jumping at a specific geographic location is calculated as the number of samples that managed to jump at this location divided by the total number of samples tested (i.e. 30 000). A map of probability of jumping, $P_{jump}$, can thus be estimated, as shown in Figure 5.a. The probability of jumping on the second fault is estimated as the maximum probability sampled at the location of the fault. In this study, this probability is 59%.

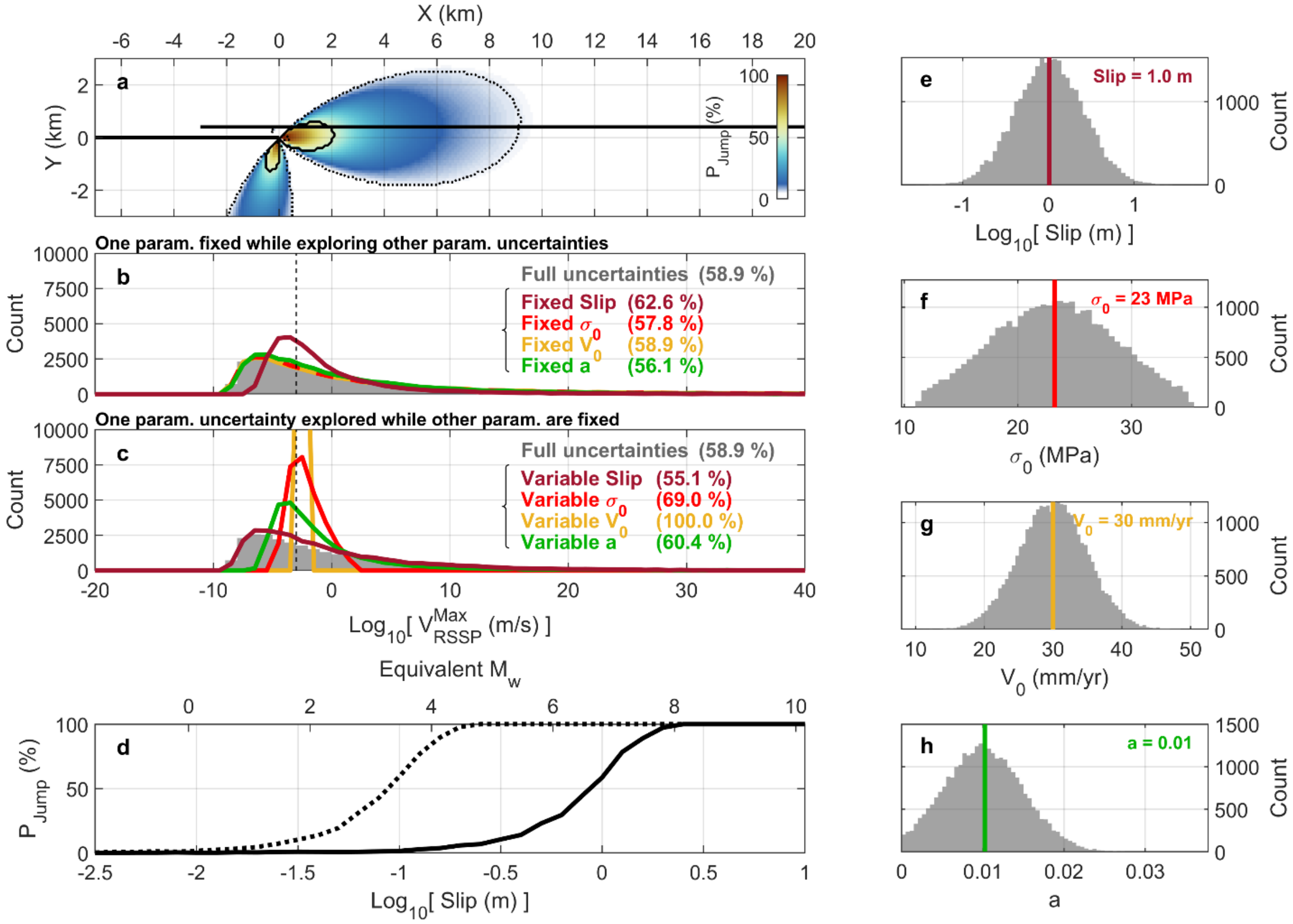

**Figure 5**: Probability $P_{Jump}$ of an event of 40 km length to jump on a second fault, fixing the angle of the receiver fault to 0°. The probabilities were computed using 30 000 samples from the distribution of the parameters indicated in (e), (f), (g) and (h). (a) Map of $P_{Jump}$. The black horizontal lines represent the source and receiver faults. Contours in full and dotted line correspond to the probabilities of 50% and 5%, respectively. The grey histogram in panel (b) and (c) is the distribution of the maximum $V_{RSSP}$ on the receiver fault computed exploring the uncertainty of all the parameters. (b) Distribution of $V_{RSSP}$ when one parameter is fixed (see values of the vertical colored lines in panel (e) to (h)) while the uncertainty of the others is explored. The vertical black dashed line represents the threshold of $10^{-3}$ m/s over which an event is assumed an earthquake and thus has jumped to the 2nd fault. The probabilities of jumping are indicated in parenthesis. (c) Distribution of $V_{RSSP}$ when the uncertainty of one parameter is explored while the other parameters are fixed. (d) $P_{Jump}$ as a function of slip for a distribution of $\sigma_0$ centered around 23 MPa (full black line) and 2.3 MPa (dotted black line).

We can also retrieve for each of the 30 000 tests the distribution of $V_{RSSP}$ along the second fault and select for each distribution the maximum value. The histogram of maximum $V_{RSSP}$ at the location of Fault 2 is shown in grey in Figure 5.b and c. The probability of jumping is calculated here as the number of maximum $V_{RSSP}$ above $10^{-3}$ m/s divided by the number of samples. This probability is estimated at $P_{jump}$= 59%.

We can then provide a sensitivity test by either fixing one parameter while exploring the uncertainty of other parameters (Figure 5.b) or by exploring the uncertainty of one parameter while the other parameters are fixed (Figure 5.c). For the 1st case, when $\sigma_0$, $V_0$ or $a$ are fixed the shape of the maximum $V_{RSSP}$ distribution does not change. On the contrary, when fixing only the slip the maximum $V_{RSSP}$ distribution changes (Figure 5.b) which suggests that the slip uncertainty controls the probability $P_{jump}$. This is confirmed when exploring the uncertainty of one parameter while the other ones are fixed (Figure 5.c). Exploring the uncertainty of slip provides a distribution of maximum $V_{RSSP}$ similar to when the uncertainties of all the parameters are explored. The distribution of maximum $V_{RSSP}$ is sharp when exploring the uncertainty of $V_0$ (-3.3 and -3.0 $log_{10}(m)$ for 1 and 99 percentiles, respectively), highlighting the weak weight of this parameter in the calculation of the final probability, given the uncertainties we determined. Finally, we see that the distribution of maximum $V_{RSSP}$ when exploring the uncertainty of $a$ (and to a lesser extent $\sigma_0$) has a shape similar to the one when the slip is the only parameter fixed (brown curve in Figure 5.b). This shows that the uncertainty of $a$ is the second most important parameter for the calculation of the probabilities, given the uncertainties explored here. Diminishing the uncertainty of the length-slip scaling law would give more weight to the uncertainty of $a$ as shown in Figure S8. This sensitivity test highlights here the weight of the uncertainties of each parameter, uncertainties that could be diminished by future studies or site specific data.

Since slip mainly controls $P_{jump}$, we show in Figure 5.d how $P_{jump}$ evolves with slip. In the example in the paragraph above, $\sigma_0$ and related uncertainties are determined for a case of pseudo-depth of 1 km. In reality, the depth at which earthquakes jump is not well documented and constrained, and $\sigma_0$ depends also on other parameters than just depth, including pore pressure. In Figure 5.d we show additionally the results for a pseudo-depth of 0.1 km, with $\sigma_0$ following a normal distribution of $N(2.3,0.6)\ MPa$. For the 1 km pseudo-depth case (~23 MPa), we observe that $P_{jump}$ starts to increase smoothly at $\sim log_{10}(slip) = -1.0\ log_{10}(m)$ (i.e. 0.1 m) before reaching almost a 100% at $log_{10}(slip) = 0.3\ log_{10}(m)$ (i.e. 0.5 m). For the 0.1 km pseudo-depth case (~2.3 MPa), everything is shifted $-1.0\ log_{10}(m)$ towards lower slip values. The shape of $P_{jump}$ is controlled here by the uncertainties of the other parameters, mainly $a$ and $\sigma_0$.

## 6.2. Examples with angle uncertainties between two faults

Until now the angle between Fault 1 and 2 was fixed. We provide here examples on how to take into account the uncertainty of the angle between the two faults.

Changing the angle between the two faults will change the pattern of Coulomb and normal stress change (Figure S4 and S5) as well as $\sigma_0$ which depends on the angles of the faults relative to the regional principal static stress field. We create first an abacus of the Coulomb and normal stress changes for every 5° interval, stress changes which are normalized by the slip for the same reason as explained in Section 5.1, i.e., the hypothesis on the uniform slip distribution for seismic events and the negligible impact of the stress field due to the second tip of the source fault. This abacus makes the calculations faster. The regional static stress field is fixed at the optimal angle of 30° relative to the source fault. We explore here the uncertainty of the normal stress, $\sigma_0$, of the source fault similarly to the previous section (i.e. $N(23.2,6.0)\ MPa$), calculate the principle stresses $\sigma_1$ and $\sigma_3$ based on the $\sigma_0$ sampled assuming that the Mohr circle should be tangent to the Mohr criteria using a coefficient of friction of 0.6, and sample the normal stress on the receiver fault using the appropriate angle relative to $\sigma_1$ (see Figure S9).

Based on those uncertainties, we suggest here two approaches to include the effect of the angle uncertainty between the two faults. The first assumes the patches of the receiver fault have an angle uncertainty that follows a Gaussian distribution, here centered on 0° with 10° as standard deviation. It assumes that even though the position of the fault is known, the sub-patches of the fault have a roughness following the chosen distribution. While we chose here a Gaussian distribution for simplicity, any other type of distribution can be chosen as input (e.g. Brodsky et al., 2016; Candela et al., 2011). The results are shown in Figure 6.a. As it is difficult to distinguish any differences between Figure 5.a and 6.a, we show in Figure 6.b the difference between the two maps (probability map with angle uncertainty centered on 0° minus the probability map for a fixed angle of 0°). We observe that probabilities drop slightly (~10%) at the position of the probability distribution for a fixed angle while it increases slightly at its borders (~15%). Indeed, some samples from the 30 000 possible have now angles closer to 10 or 20°, and thus give less weight in terms of probability for angles at 0°. The second approach to include angle uncertainties assumes that there will always be a fault patch optimally oriented for an earthquake to jump. Effectively, for each 30 000 samples, all angles are tested, and for each geographic location, the angle which produces the maximum $V_{RSSP}$ is selected. The results are shown in Figure 6.c and show two main high probability lobes at an extensional position. A map of the optimal angle, considering the uncertainties explored, is also presented in Figure S10. Note that the extensional position is also favored here as it is located parallel to the maximum principal stress $\sigma_1$ which induces minimum normal stress ($\sigma_0$ equal to $\sigma_3$) and maximizes $V_{RSSP}$. The angle of the principal stresses is, to some extent, a modifier of the probability map without the effect of the regional stress (i.e. taking the same normal stress for both faults). An example of this second approach without the effect of the regional stress is shown in Figure S11 and still highlights the two main high probability lobes at an extensional position but with less force.

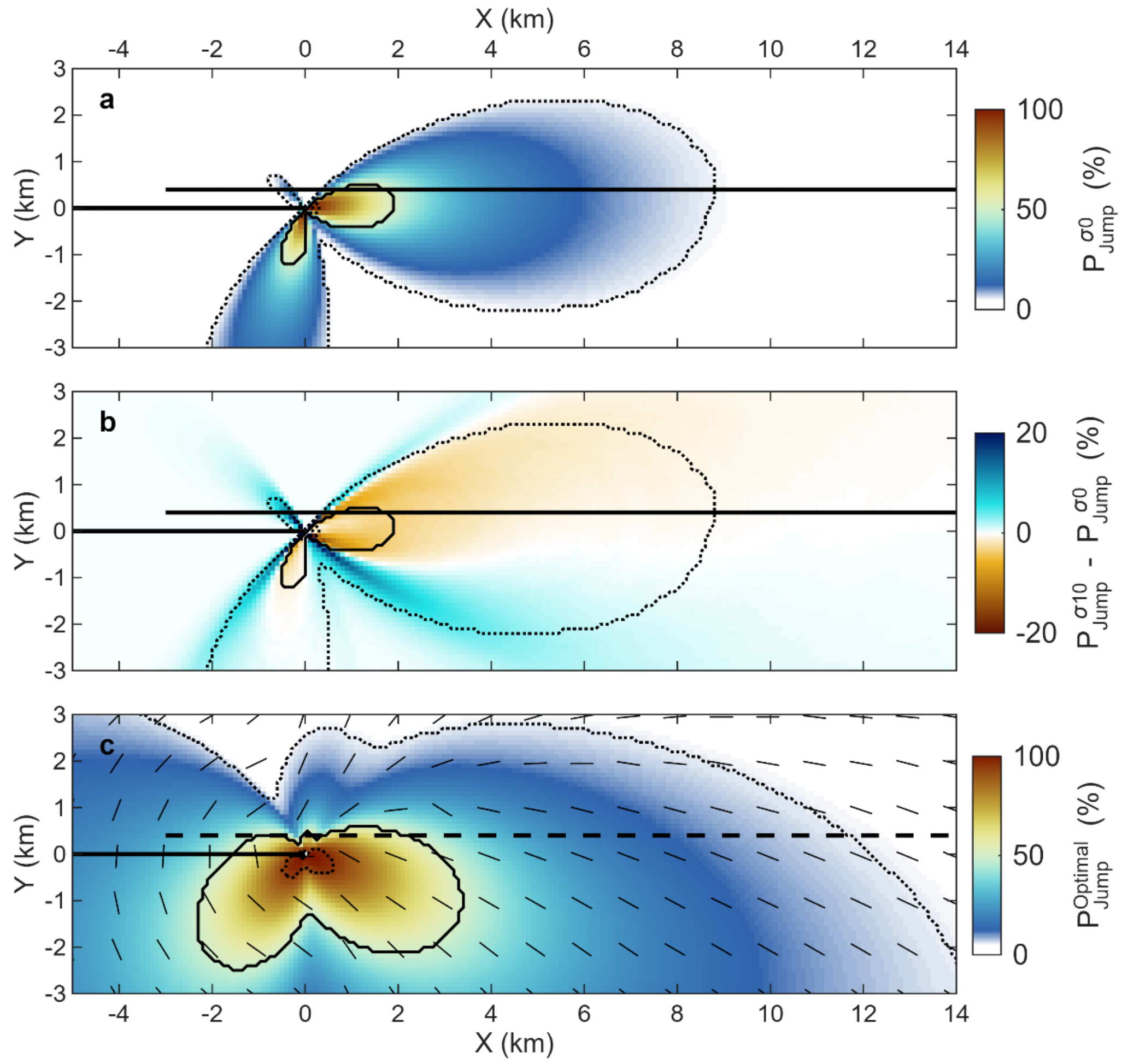

**Figure 6**: Impact on $P_{Jump}$ of exploring the uncertainty of the angle between the source and receiver fault. (a) Same as Figure 5.a but with an angle uncertainty which follows a Gaussian distribution centered on 0° and with a standard deviation of 10°. (b) Difference between the map of $P_{Jump}$ while exploring the angle uncertainty (Figure 6.a) and the one where the angle is fixed (Figure 5.a). (c) Map of $P_{Jump}$ assuming that there will always be a fault patch optimally oriented for an earthquake to jump on the receiver fault (represented by the black dashed line). The small thin black lines indicate the angle at which the receiver fault is optimally oriented to host a jump.

# 7. Discussion

In this study, we highlight the key factors controlling an earthquake jumping from one fault to another, based on the rate-and-state friction law and the geometrical configuration of two faults. The angle between the two faults controls the pattern of high and low probability through the normalized Coulomb ($\overline{\Delta C}$) and normal ($\overline{\Delta \sigma}$) stress changes (see eq. [6]). The amplitude of this pattern is then modulated by the slip, $\sigma_0, a, \text{and } V_0$. While we can manage to provide rough uncertainties for the slip, $a, \text{and } V_0$, values for the parameter $\sigma_0$ are trickier to impose. This is important as very low value of $\sigma_0$ lead to potentially very large jump distances, up to infinity if $\sigma_0$ approaches 0.

We provide here an illustration of an approach to constrain the values of $\sigma_0$ using the empirical probabilities from Biasi & Wesnousky (2016). In their dataset of 76 earthquakes, including 46 in strike-slip regime, no events managed to pass a step greater than or equal to 6 km, which results in a probability to jump such steps equal to 0. Among those events, the 2010 Yushu earthquake and its foreshock of magnitude $M_w$6.8 and 6.1, respectively, failed to pass a step of 6 km in an extensional regime. We test here the minimum $\sigma_0$ possible for such events to fail to jump a 6 km step. To do so, we explore values of $V_0$ using the normal distribution $N(3.5\ ,0.5)\ mm/yr$ since the slip rate on the Yushu segment is about 3-4 mm/yr (Zhang et al., 2022), and explore the same uncertainties for $a$ as in the previous Sections. We calculate for a setting of two overlapping parallel faults, for both restraining and extensional regimes, the probability $P_{jump}$ as a function of $\sigma_0$ for different values of slip (Figure 7.a and b). The 5, 50 and 95 percentiles of those curve are estimated and reported in a slip versus $\sigma_0$ map (Figure 7.c et d). According to a co-seismic slip inversion based on GPS and InSAR data (Wen et al., 2013), the maximum slip reaches 2 m while surrounding areas slip roughly one order of magnitude less (~0.2 m). For a step of 6 km in extensional regime, as for the Yushu earthquake, the minimum $\sigma_0$ possible for a slip of 0.2 m is 2.1 MPa (120 m in pseudo-depth) taking the 5% percentile of $P_{jump}$ (Figure 7.d). Note that while this example takes advantage of earthquakes that failed to pass fault steps to retrieve a lower bound of $\sigma_0$, one can also use earthquakes that succeeded to pass to constrain an upper bound of $\sigma_0$.

As $\sigma_0$ decreases together with depth, the probability to jump to secondary faults at further distance increases (Figure 4.a and c), which increases slip distribution and partitioning at the surface. It supports the idea that the total displacement partitioned at the surface along secondary faults (off-fault deformation) is probably equivalent to the displacement occurring at depth on the main rupture (Antoine et al., 2024).

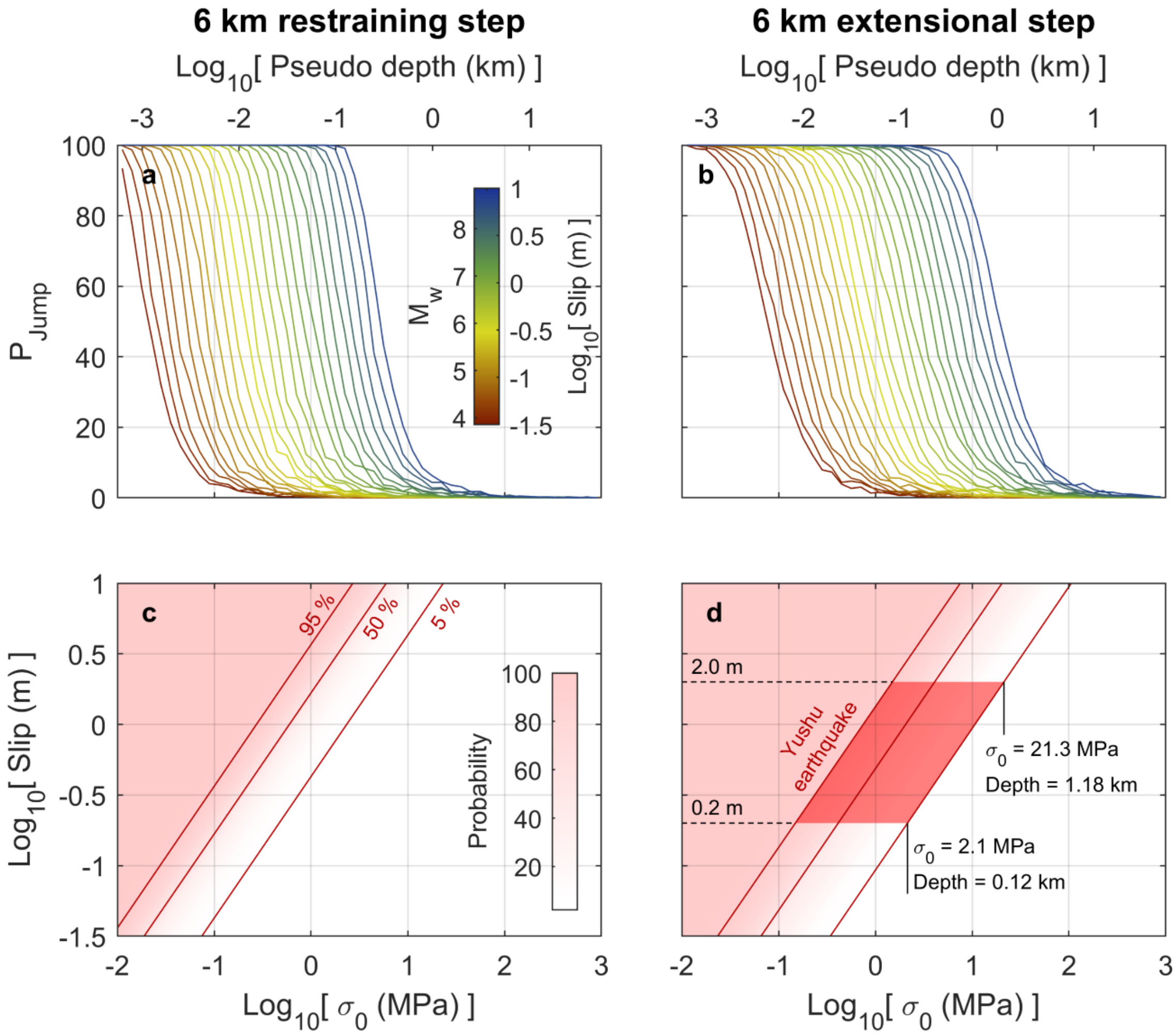

**Figure 7**: $P_{Jump}$ as a function of slip and $\sigma_0$. (a) $P_{Jump}$ as a function of $\sigma_0$ for different values of slip and for a restraining step of 6 km. (b) Same as (a) but for an extensional step. (c) Map of $P_{Jump}$ in the slip-$\sigma_0$ space for a restraining step 6 km. The dark red lines indicate the position of 5, 50 and 95 % probabilities. (d) Same as (c) but for a extensional step. The red patch indicates the values of slip and $\sigma_0$ within 5 and 95% probability associated with the 2010 Yushu earthquake and its foreshock of magnitude $M_w$6.8 and 6.1, respectively, that failed to jump a 6 km step. According to this diagram, the Yushu earthquakes fails to jump at probabilities above 95% ($P_{Jump}$ <5%) for $\sigma_0$ above 2.1 MPa (i.e. pseudo depth>120 m), assuming slip of 0.2 m.

The calculations of probabilities in Section 5 expected earthquakes to stop at the tip of the source fault. But an earthquake can stop before reaching it as a result of specific stress conditions on the fault (Michel et al., 2017). To take into account this aspect, it might be more reasonable to extrapolate along the source fault the highest probabilities to jump (Figure S12).

The probability calculations in Section 5 did not take into account the effect of the fault tip the furthest away from the step. This effect is not negligible but decreases with the length of an earthquake as implied by the length-slip scaling laws (e.g. $log_{10}(Slip) = 0.833 * log_{10}(L) - 3.84$ for 3.4≤$L$≤45 km; Leonard, 2010). For uniform slip distributions, it is straight forward to add the second tip effect by superposing the Coulomb and normal stress map of the second tip which has an inverted pattern to the tip closest to the step. Figure S3 illustrates the impact of the second tip for uniform slip distributions.

Among the main limitations in the computation of fault jump probabilities is that the propagation and impact of seismic waves is not taken into account. The waves generated by an earthquake can pass through the receiver fault and change its shear and normal stress, and dynamically trigger a jump (Brodsky and van der Elst, 2014; Harris et al., 1991). This is additionally complexified as waves carry the source radiation pattern and directivity of the source fault, might have constructive or destructive patterns of stress change on the receiver fault and will also interact with the earth surface (Harris and Day, 1999; Hu et al., 2016; Oglesby et al., 1998). Eq. [6] might still hold considering the hypothesis behind it, but taking into account the effect of waves is challenging. Additionally, we assume here a fully elastic medium while fault damage and plastic processes actually occur during an earthquake and will modify to a certain extent the jump probabilities. Incorporating elastoplastic bulk response into seismic cycle simulations – using pressure-sensitive Drucker-Prager plasticity – have shown to modify fault behavior, especially for restraining steps (Figure S7), inducing rupture segmentation, temporal clustering, and more frequent jumps from one fault to another (Mia et al., 2024). The framework accounts for ruptures that jump immediately from one fault to another, but does not capture ruptures that occur with a slight delay after the initial rupture on the source fault. Note that the $V_{RSSP}$ criterion becomes less accurate as slip rate increases, as discussed in Text S6. The definition of an earthquake, defined here using a slip rate detection threshold of $10^{-3}$ m/s, has thus implications but does not affect the conclusions of our study (Text S6). We interpret that once the fault reaches this detection threshold, rupture is very likely to transition to dynamic behavior, as observed in most simulations. Additionally, although the simulations show a temporal evolution of $V_0$ (e.g. the 'healing' period; see Section 3 and Figure S2.c), the probabilities calculated with the criterion are not time dependent. We assume that the $V_0$ of the receiver fault is always at the loading rate.

Finally, note that the predictions of our approach in terms of fault jump are coherent with the results of other studies using rate-and-state friction (e.g. Kroll et al., 2023; see Text S7 for details).

## 8. Conclusion and Perspective

This study focuses on characterizing a criterion to evaluate the probability of an earthquake to jump from one fault to another. This criterion (Eq. [6]), $V_{RSSP}$, is based on the rate-and-state friction law and assumes an instantaneous stress interaction between the source fault, hosting the earthquake, and the receiver fault, on which the earthquake rupture might propagate. 2D quasi-dynamic seismic cycle simulations were used to confirm the validity of the criterion in the rate-and-state framework. We further proposed an approach to evaluate the probability of an earthquake to jump and provided a sensitivity test of $V_{RSSP}$. 2D settings were presented as examples in this study, but the approach used here can be applied to 3D problems.

The criterion evaluates the initiation of seismic slip on the receiver fault but does not determine whether the rupture will propagate further. The continuation of the rupture propagation depends on the state of the receiver fault, for which the amount of elastic energy accumulated since the previous earthquake likely plays an important role (Weng and Ampuero, 2019, 2020).

The criterion depends on parameters that can be potentially estimated or measured. $\Delta C$ and $\Delta\sigma$ regroup the information on both the earthquake slip distribution of the source fault and on the geometry of the step. But they are insufficient by themselves to predict if an earthquake will jump or not. $a$ can be estimated from experimental studies (e.g. Blanpied et al., 1991; Cappa et al., 2019). $V_0$ can be assumed close to the source fault loading rate. Estimating $\sigma_0$ concentrates the main challenges as small values implies longer jump distances, distances which seem improbable considering observations (Biasi and Wesnousky, 2016). To constrain the probabilities of fault jump, it will be necessary to characterize the depth profile of $\sigma_0$ in the very shallow region, to evaluate if a minimum $\sigma_0$ exists, to understand if earthquake slip deficit, slip partitioning, effect of free surface and plastic processes among other phenomenas at the surface counteract the effect of the decrease of $\sigma_0$ on jump distances. Finally, it is important, using the data available (e.g. seismological records), to observe and document more thoroughly and systematically the location at which fault rupture jump occurs, whether it happens in the deeper or shallower portion of the fault.

## Open Research

All the necessary data to reproduce the results are available at (Michel et al., 2025).

## Conflict of Interest

The authors declare no conflicts of interest relevant to this study.

## Acknowledgements

This project has received funding from the Agence National de la Recherche (ANR EQ-TIME; Projet-ANR-19-CE31-0031). PR acknowledges support the European Research Council (ERC) Starting Grant 101040600 (HYQUAKE). A significative portion of the analyses reported in this paper were done using MATLAB (The MathWorks Inc., 2023). We thank the reviewers and editors for their insightful comments which helped improve the study substantially.

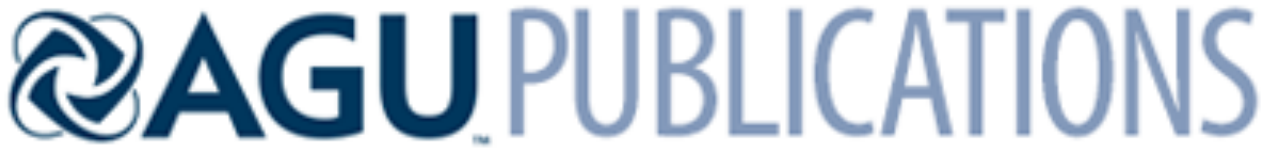

**Probability of earthquake fault jumps from physics based criterion.**

Sylvain Michel[1,2,3,4], Oona Scotti[1,5], Sebastien Hok[1,5], Harsha S. Bhat[4], Navid Kheirdast[4], Pierre Romanet[6,2], Michelle Almakari[4], Jinhui Cheng[4, 7]

[1] Institut de Radioprotection et de sûreté Nucléaire, 31 avenue de la Division-Leclerc, 92262, Fontenay-aux-Roses, France

[2] Université Côte d'Azur, CNRS, IRD, Observatoire de la Côte d'Azur, Géoazur, Sophia-Antipolis, France

[3] CNRS-INSU, Institut des Sciences de la Terre Paris, Sorbonne Université, ISTeP UMR 7193, 75005 Paris, France

[4] Laboratoire de Géologie, Département de Géosciences, Ecole Normale Supérieure, PSL Université, CNRS UMR 8538, 24 Rue Lhomond, 75005, Paris, France.

[5] Now at Autorité de Sûreté Nucléaire et Radioprotection, 31 avenue de la Division-Leclerc, 92262, Fontenay-aux-Roses, France

[6] Department of Earth Sciences, La Sapienza University of Rome, Rome, Italy

[7] Now at Division of Geological and Planetary Sciences, California Institute of Technology, Pasadena, USA

**Contents of this file**

**Introduction**

This supplementary material provides information about (1) the difference between the criterion from Kroll et al. (2023) and the one from this study, (2) the use of different loading rates in the simulations, (3) the exploration of the geometrical parameter H of the simulations, (4) the selection procedure of events on Fault 1 with high Coulomb stress change but no jump,

(5) the determination of the normal stress gradient with depth and its uncertainties, (6) the implication of the definition of an earthquake on the results and the criterion, and (7) a comparison with the study from Kroll et al. (2023). It also contains additional figures which illustrate further the content in the main text.

**Text S1. Difference between the criterion from Kroll et al. (2023) and the one from this study**

In Kroll et al. (2023), the criterion takes the following form: $V_{Kroll} = V_0 \, exp\left[ \frac{1}{a\,\sigma_0} \left( \Delta\tau - \frac{\tau_0}{\sigma_0} \Delta\sigma \right) \right]$.

$\Delta\sigma$ does not appear in the denominator within the exponential (see Equation [6]). While this assumption may hold at greater depths—provided fluids do not significantly reduce the effective normal stress—closer to the Earth's surface, $\Delta\sigma$ is likely to become more significant.

**Text S2. On the use of different loading rates in the simulations.**

Using the same loading rate for both faults tends to synchronize their behavior: earthquakes on Fault 1 occur during the same phase of Fault 2's interseismic period. To avoid such bias, we aimed for earthquakes on Fault 1 to interact with Fault 2 at various stages of its seismic cycle, thereby providing a wide range of scenarios. The most straightforward way to achieve this was to introduce a contrast in the loading rate between the two faults. Varying the loading rate contrast affects the frequency at which earthquakes on Fault 1 attempts to jump on Fault 2. If the contrast is too small, some synchronization between the two faults might occur.

**Text S3. On the exploration of the geometrical parameter H of the simulations.**

H is the perpendicular distance between the closest tips of the two faults, taken relative to the direction of the first fault (Figure 1). The range of H explored in this study is limited (Table 1). But it reflects the importance of the impact of the normal stress on fault jumps considering the rate-and-state framework. In the simulations, the normal stress applied to the faults are 80 and 40 MPa for the 1st and 2nd scenarios, respectively. For jumps to occur, and sometimes fail, at this level of normal stress, H needs to be quite small (i.e. of the order of what has been tested) or the earthquake on the source fault quite large, otherwise they will always fail. For scenario 3, with a normal stress around 80 MPa, it is very difficult to jump very far. While the simulations have been set up with a limited range for H, the criterion and our study is aimed for a general case, i.e. steps as large as wanted. The simulations are only here to test the criterion $V_{RSSP}$.

**Text S4. Selection of events on Fault 1 with high Coulomb stress change but no jump.**

For each scenario we select events that did not jump from Fault 1 to 2, but would have if their $V_0$ equaled the loading rate (green dots in Figure 3). In those circumstances, even though an event has a high coulomb and normal stress change, they do not necessarily jump. Figure S13 provides clearer examples of this class of events than Figure 3.

**Text S5. Determination of the normal stress gradient and its uncertainties**

In our study, the gradient of the normal stress with depth is determined as follow. We set ourselves on a fault in strike-slip regime. The maximum and minimum principal stresses, respectively $\sigma_1$ and $\sigma_3$, are thus both horizontal, and the vertical principal stress, $\sigma_V$, corresponds to $\sigma_2$, such that $\sigma_3 \leq \sigma_V \leq \sigma_1$. We take a gradient of $\sigma_V$ with depth of 23 MPa/km as suggested by Zoback et al. (2003) for clastic sedimentary rocks and hydrostatic conditions. We also assume that the fault is optimally oriented (i.e. $\sigma_1$ is 30° from the fault taking a coefficient of friction of 0.6) and that the Mohr circle calculated from the principal stresses is tangent to the Mohr-Coulomb failure envelope. Based on the gradient of $\sigma_V$, the minimum and maximum values for $\sigma_1$ and $\sigma_3$ are calculated when $\sigma_1 = \sigma_V$ and $\sigma_3 = \sigma_V$, respectively. The minimum and maximum normal stress, $\sigma_0$, on the fault at 1 km depth is thus equal to 11.3 and 35.2 MPa (vertical blue and orange dashed lines in Figure S9.a). We take here an average value of the minimum and maximum gradient, which equals to 23.2 MPa/km (vertical green dashed line in Figure S9.a). The pseudo-depths shown in Figure 4 and 7 uses this gradient.

The values and uncertainties of $\sigma_0$ taken in Section 4 (Figure 5.f) are set as follow. We assume that the distribution of the normal stress on the generator fault is Gaussian, centered on the gradient determined in the previous paragraph (i.e. 23.2 MPa/km). The two standard deviation extent of the Gaussian is set to the minimum and maximum normal stress gradient (i.e. 11.3 and 35.2 MPa/km). The normal distribution of the generator fault is truncated at 11.3 and 35.2 MPa/km so that the scenario is constrained to a strike-slip regime. The normal stress on the receiver fault, $\sigma_0$, thus depends on the principal stress conditions set on the generator fault and on the receiver fault orientation relative to $\sigma_1$.

**Text S6. On the implication of the definition of an earthquake.**

The timing and spatial extent of earthquakes are defined here using a slip rate detection threshold of $10^{-3}$ m/s, following Romanet et al. (2018), which introduced the algorithm VEGA – the quasi-dynamic numerical simulator of seismic cycle used in this study. This threshold is not standardized across the literature. For example, some studies adopt a higher value of $10^{-1}$ m/s (e.g. Lapusta & Liu, 2009). For comparison, we present simulation results obtained using a slip rate thresholds of $10^{-2}$ m/s in Figure S13. The main conclusions of our study remains unchanged.

Nevertheless, increasing this threshold have some implications. The accuracy of $V_{RSSP}$ relies on the assumption that the state variable in the rate-and-state friction law does not have time to evolve. At sufficiently high slip rates, this assumption breaks down, as illustrated by the aging law used in this study (Ruina, 1983): $\frac{d\theta}{dt} = 1 - \frac{V\theta}{L}$. As slip rate $V$ increases, $\left|\frac{d\theta}{dt}\right|$ also increases, indicating more rapid evolution of the state variable. Consequently, the $V_{RSSP}$ criterion is likely to become less accurate at high slip rates. Importantly, this limitation does not affect the conclusions of our study. Our interpretation is that once the fault reaches the chosen detection threshold, rupture will transition to dynamic behavior.

We examine the temporal evolution of $\left|\frac{d\theta}{dt}\right|$ as a function of $V$ in one simulation of scenario 1 on Fault 2. A change of $\left|\frac{d\theta}{dt}\right|$ is defined as *small* when it is less than a fraction $\chi$ of $\theta$. In this analysis, we set $\chi = 5\%$. Figure S14.a shows the evolution of $|\chi\theta| - \left|\frac{d\theta}{dt}\right|$ over the entire simulation for all velocity weakening subfaults. Small changes in $\left|\frac{d\theta}{dt}\right|$, according to our definition, occur when $|\chi\theta| - \left|\frac{d\theta}{dt}\right| > 0$ (red dots in Figure S14.a), and correspond mostly when $V$ is below ~$10^{-3}$ m/s, although some time steps with $|\chi\theta| - \left|\frac{d\theta}{dt}\right| < 0$ (black dots inf Figure S14.a) are still observed down to $V$~$10^{-5}$ m/s. However, when restricting the analysis to the periods near the onset of seismic events — from half a year before event initiation to a tenth of the total event duration — large changes of $\left|\frac{d\theta}{dt}\right|$ only occurs for slip rates exceeding ~$10^{-3}$ m/s (Figure S14.b). This result illustrate the range of validity of the criterion within the simulations and supports the choice of $10^{-3}$ m/s as a reasonable velocity threshold for defining earthquakes in this modeling framework.

**Text S7. Comparison with the study from Kroll et al. (2023)**

Kroll et al. (2023) test whether a rupture is able to jump from one fault to another, in a step geometry, using two models: the quasi-dynamic model RSQSIM and fully dynamic FaultMod. For both cases they simulate single ruptures on planar 3D faults reaching 20 km depth but assuming a constant initial normal and shear stress over the whole fault (60 MPa and 28.38 MPa, respectively). Using Equation [6] from our study, we retrieve coherent results with Kroll et al. (2023) using their parametrization: $V_{theory} \approx 0.15\ m/s$. For this test, we assumed $a = 0.01$, $\sigma_0 = 60MPa$, and $\Delta C = 1$ MPa, which is a rough value according to the Coulomb stress change value at the renucleation locations in figure 5 of their paper. As $V_0$ is not available in their study, we fixed it at 30 mm/yr. We also chose a normal stress change, $\Delta\sigma$, of 1 MPa, similar to the Coulomb stress change, but it has in this case almost no impact as the effect of $\sigma_0$ will dominate according to Equation[6] (i.e. $\sigma_0 \gg \Delta\sigma$).

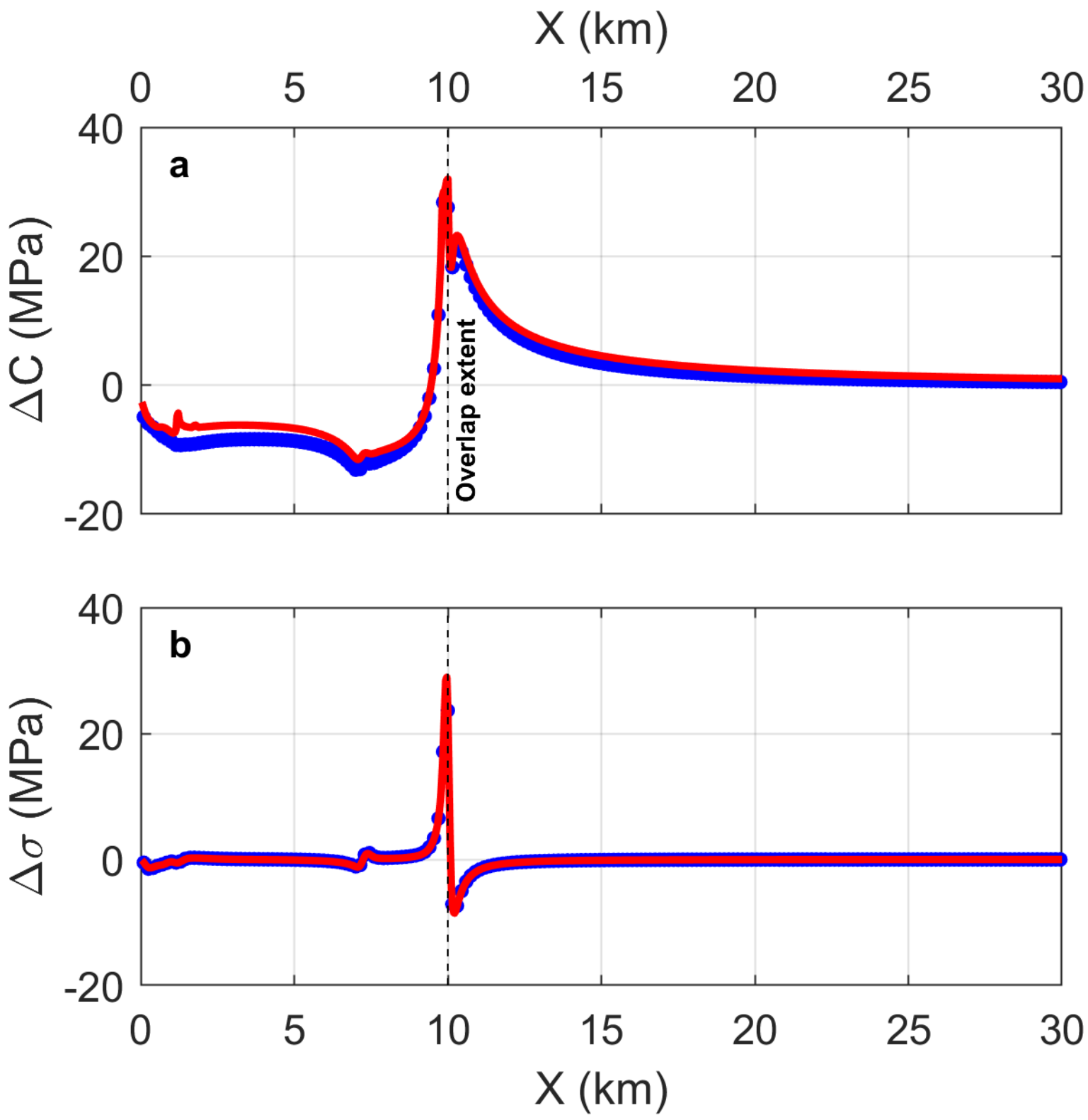

**Figure S1**: (a) Coulomb and (b) normal stress change profiles of event 6 (see Figure 2) along the receiver fault estimated from the simulation (red line) using our procedure in Section 4 and using Okada's approach based on event 6 slip distribution (blue dots). The vertical dashed black line correspond to overlap extent, i.e. the position in X of the tip of the generator fault. Okada's approach allows us to estimate the effect of the static stress impact on Fault 2 without the stress redistribution occurring on fault 2 between the timing of the samples (i.e. between $t_0$ and $t_t$; Section 4) in response to the static stress change, and present in the simulations. There are little differences between the results of both calculations which suggests that the static stress impact term is dominating the stress redistribution term.

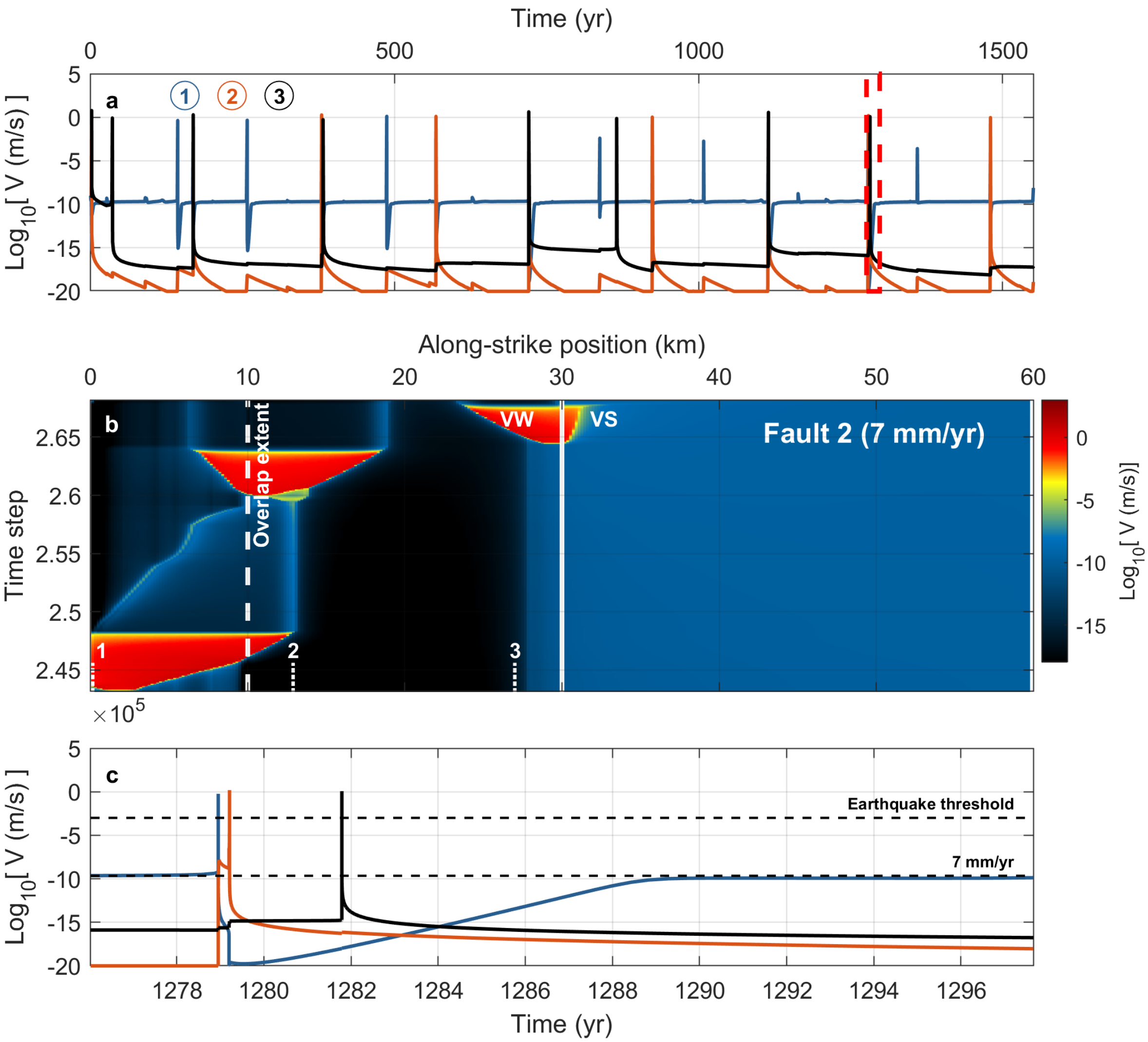

**Figure S2**: (a) Slip rate through time of three points on the fault which positions are indicated in panel (b). (b) Slip rate of Fault 2 through time. The time is indicated here in time step. In the simulations, time step size decreases when slip rate increases, which helps visualizing seismic events that last a few seconds. The position of the three points sampled on the fault are indicated by the vertical white dotted lines and numbers. VW and VS stand for Velocity Weakening and Strengthening, respectively. (c) Zoom of panel (a) as indicated by the red rectangle. Note that for point 1 (in blue), which is at the tip of the fault, the slip rate drops after an earthquake and then increases during a 'healing' period before reaching a slip rate equal to the loading rate.

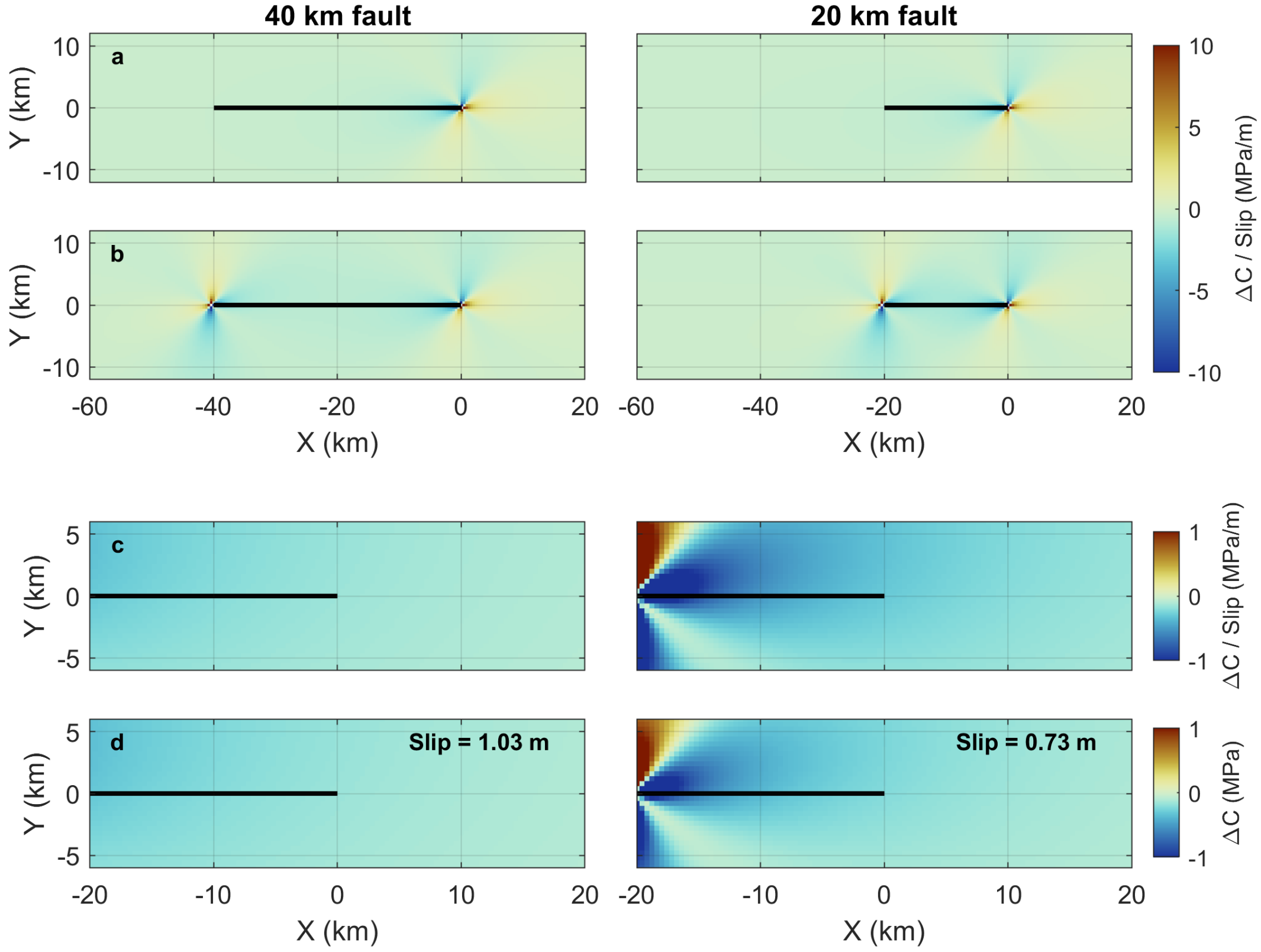

**Figure S3**: (a) shows the maps Coulomb stress change normalized by the slip on faults of 10 and 5 km length, respectively, but only taking into account the effect of one tip of the faults. Since the slip distributions are uniform and that the maps are normalized by their slip, the maps in (a) are identical. (b) is the same as (a) but now considering the two tips of the faults. The maps of normalized Coulomb stress change are consequently not identical. (c) Normalized Coulomb stress change effect of the 2$^{nd}$ tip at the location of the 1$^{st}$ tip, for the faults of 10 and 5 km length, respectively (as if the first tip had no effect). (d) Same as (c) but this time not normalized. The slips were estimated using the length-slip scaling law of Leonard, (2010). We see that the effect of the second tip is higher for a fault of 5 km than for a fault of 10 km length, considering the length-slip scaling law.

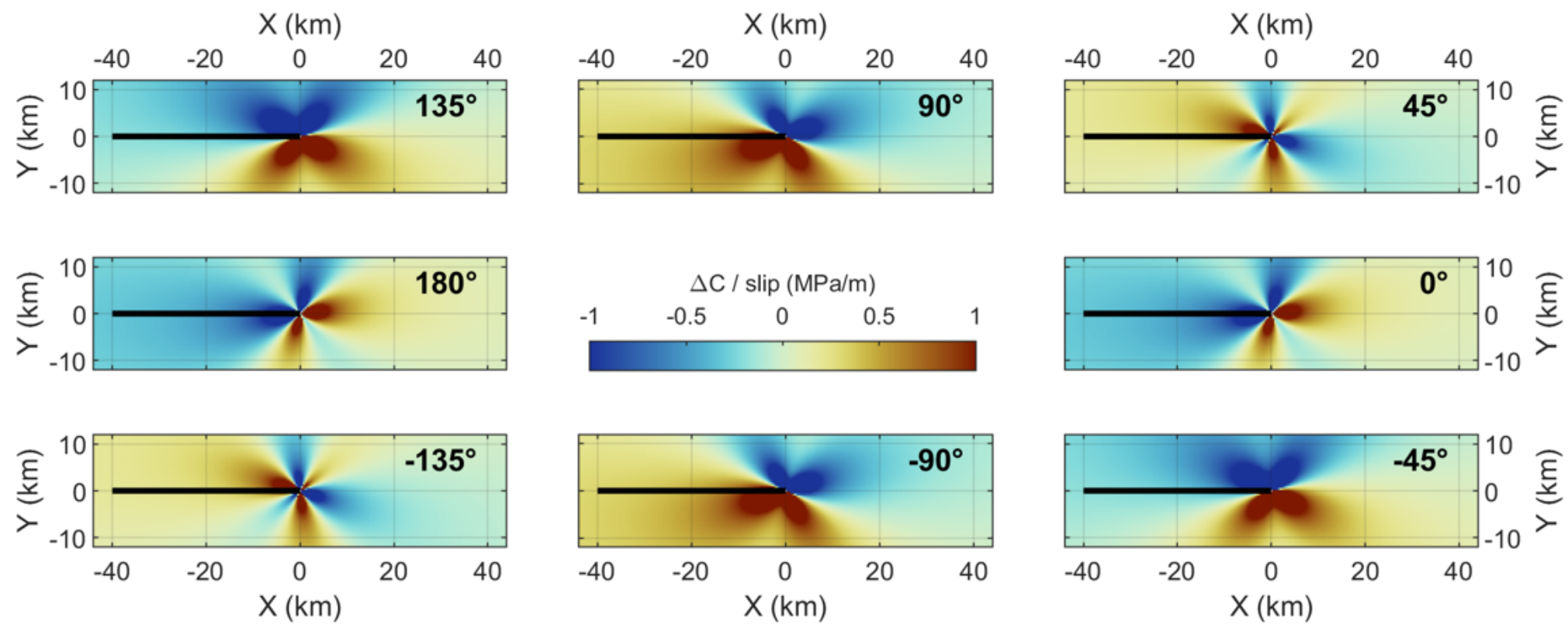

**Figure S4**: Abacus of Coulomb stress change, $\Delta C$, normalized by the slip for a uniform slip distribution and assuming no stress effect due to the second tip of the fault.

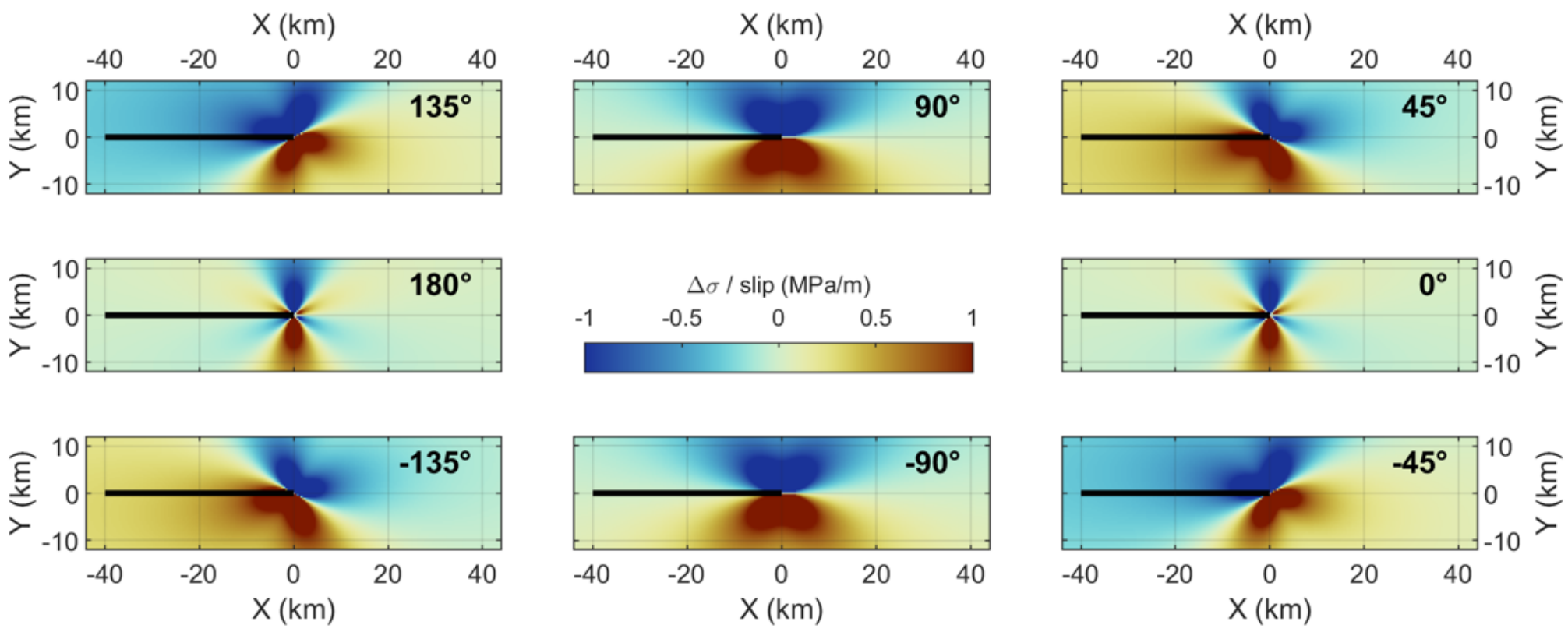

**Figure S5**: Same as Figure S4 but for the normal stress change, $\Delta\sigma$.

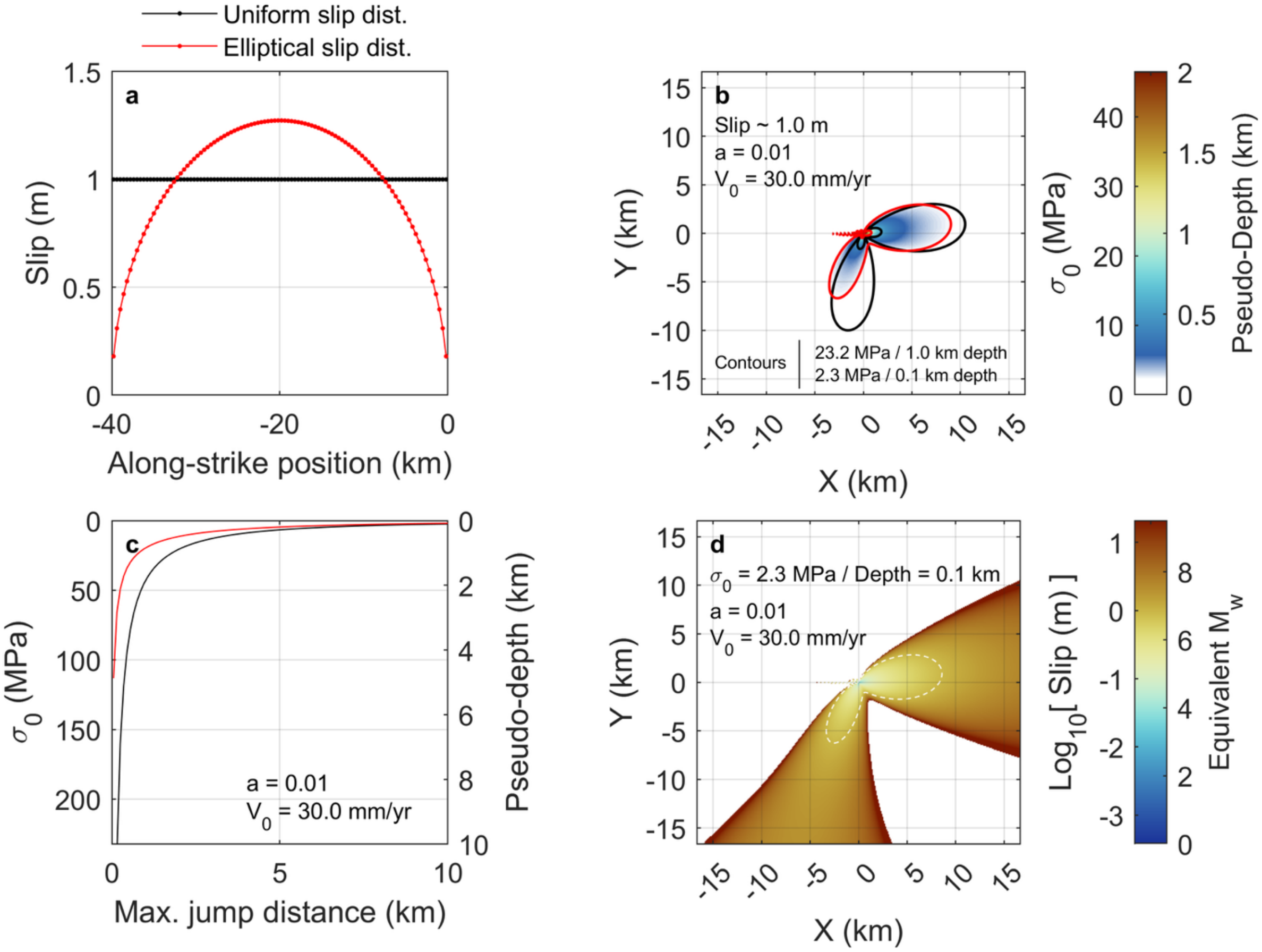

**Figure S6**: Impact on the maximum jump distance of the parameters in Eq. [6] assuming different slip distributions along the fault and no contamination of the stress impact from the second tip of the source fault, the one furthest away of a potential 2nd fault. All tests were realized using the same fixed values of $a$ and $V_0$. (a) Uniform (black) and elliptical (red) slip distribution tested. The elliptical distribution is evaluated assuming a static ('penny-shaped') crack of 20 km radius and uniform stress drop of 1.75 MPa. This elliptic slip distribution has an average slip of 1 m, similar to the uniform slip distribution. (b) Map of maximum jump distance for an associated $\sigma_0$ using the elliptical slip distribution of panel (a). The red contours indicate the position of the maximum jump distance for $\sigma_0 = 23$ and 2.3 MPa, corresponding to a pseudo depth of 1 and 0.1 km, respectively. The black contours are the same but for the uniform slip distribution in panel (a). The red profile in panel (c) is taken from this map along the coordinate Y=0. (c) Profile of $\sigma_0$ along the direction of Fault 1 as a function of the maximum jump distance, assuming the slip distributions in panel (a). $\sigma_0$ can be interpreted as a pseudo depth. We use here the gradient 23 MPa/km. (d) Map of maximum jump distance for an associated slip amplitude fixing $\sigma_0$ to 2.3 MPa (pseudo depth of ~0.1 km), based on the elliptical slip distribution. The length of the fault is here fixed to 40 km. The dashed white contour indicates the position of the maximum jump distance for the elliptical slip distribution of panel (a).

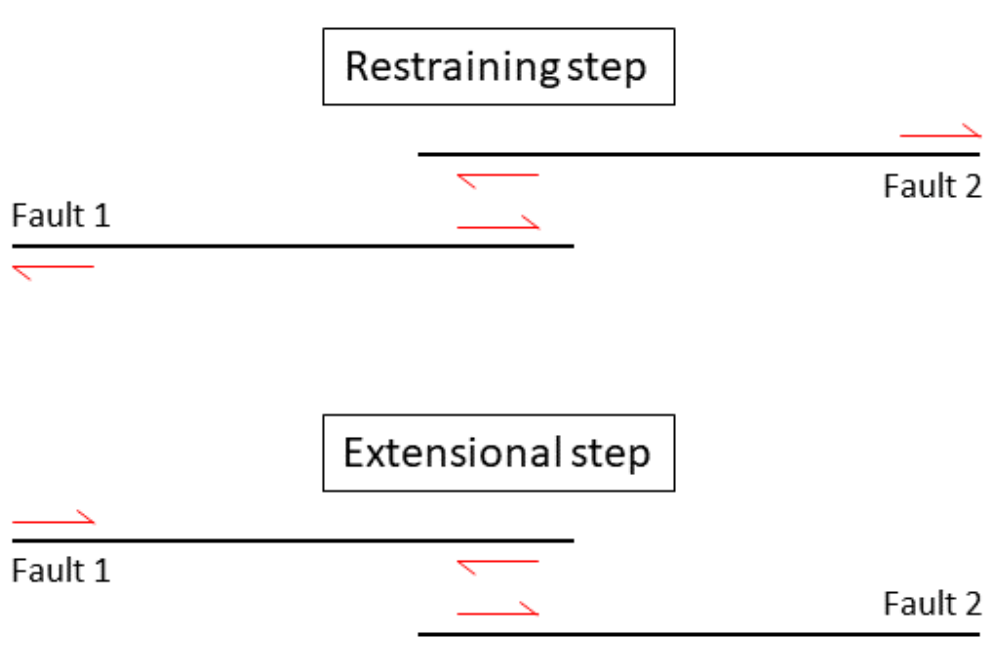

**Figure S7**: Cartoon of a restraining and extensional bends.

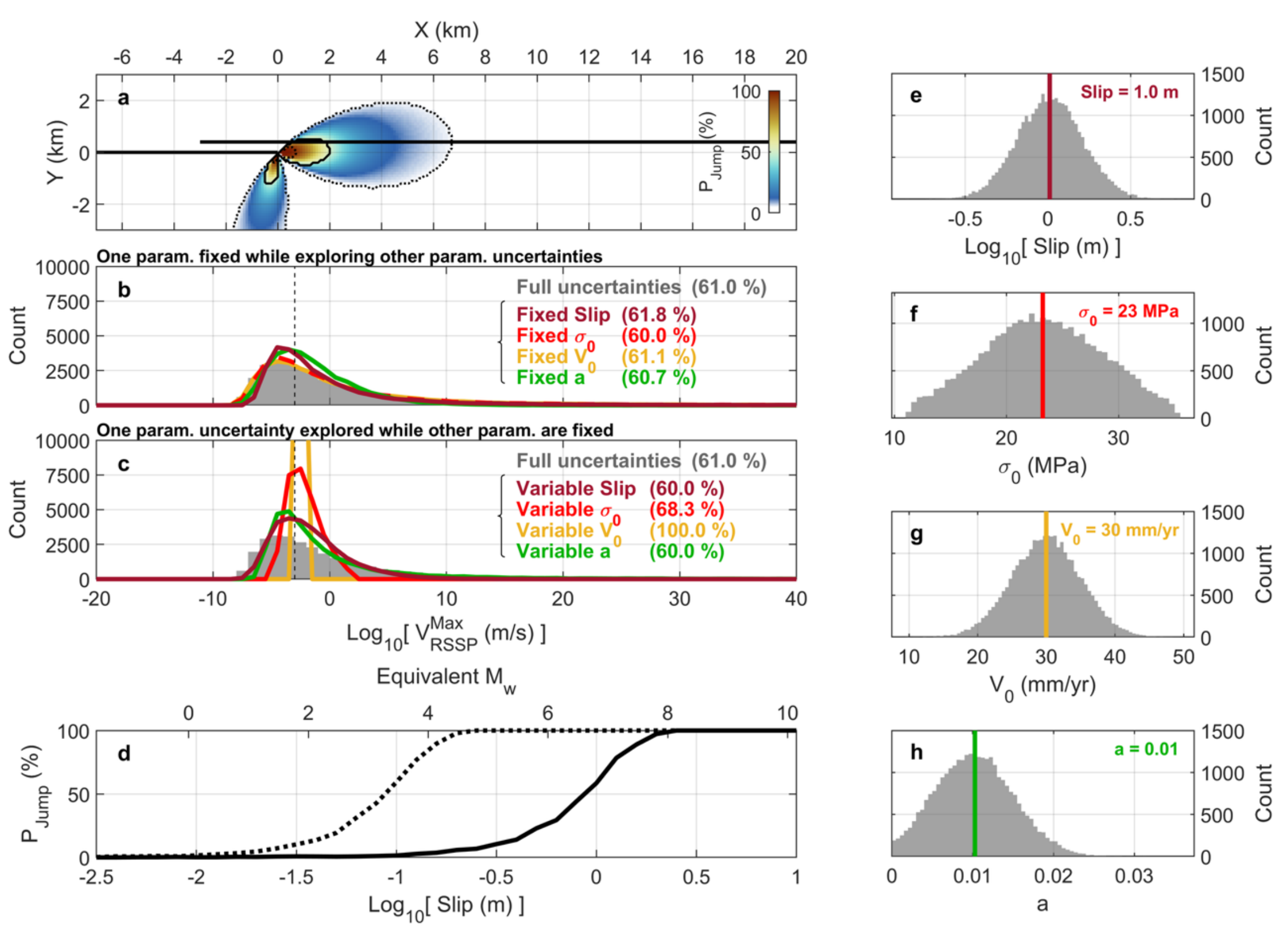

**Figure S8**: Same as Figure 5 but with the slip uncertainty divided by 2.

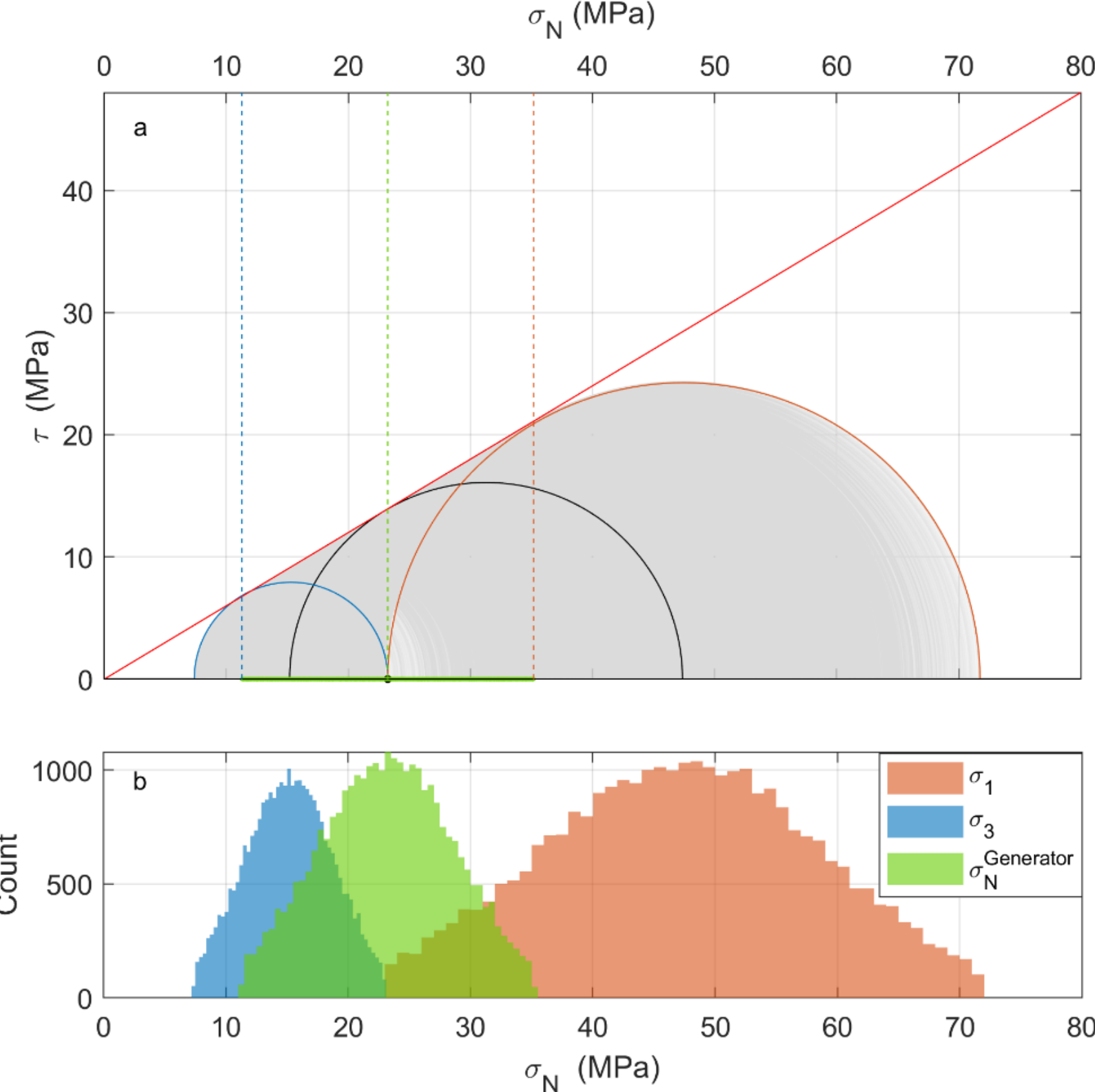

**Figure S9**: Uncertainties on $\sigma_0$. (a) Shear stress, $\tau$, as a function of normal stress, $\sigma_N$. In this study, the vertical principal stress, $\sigma_V$, is assumed equal to 23 MPa/km (see Text S1) such that $\sigma_3 \leq \sigma_V \leq \sigma_1$ since we set ourselves in a strike-slip regime. We also assume that $\sigma_1$ is also optimally oriented at 30° relative the generator fault and that the Mohr circle defined by the principle stresses is tangent to the Mohr-Coulomb failure criteria (red line). Two extreme stress conditions are thus possible with $\sigma_3 = \sigma_V$ (orange Mohr circle) and with $\sigma_1 = \sigma_V$ (blue Mohr circle). The range of normal stress on the generator fault is thus constrained between the values of normal stress, at Mohr-Coulomb failure criteria, of the two cases of extreme stress conditions (blue and orange vertical dashed lines). The distribution of the uncertainty of the normal stress on the generator fault is assumed normal, centered at the middle of the normal stress range (vertical dashed green line) with two standard deviation set to the range extent (blue and orange vertical dashed lines). All the Mohr circles explored for the generator fault are represented by the grey semi-circles, while the black circle correspond to the average one. The green dots on the X-axis correspond to the values of normal stress explored on the generator fault. (b) Uncertainty of the normal and horizontal principal stresses on the generator fault. The distributions are cut at their two standard deviation limits. The normal stress on the receiver fault depends on the orientation of the fault relative to $\sigma_1$.

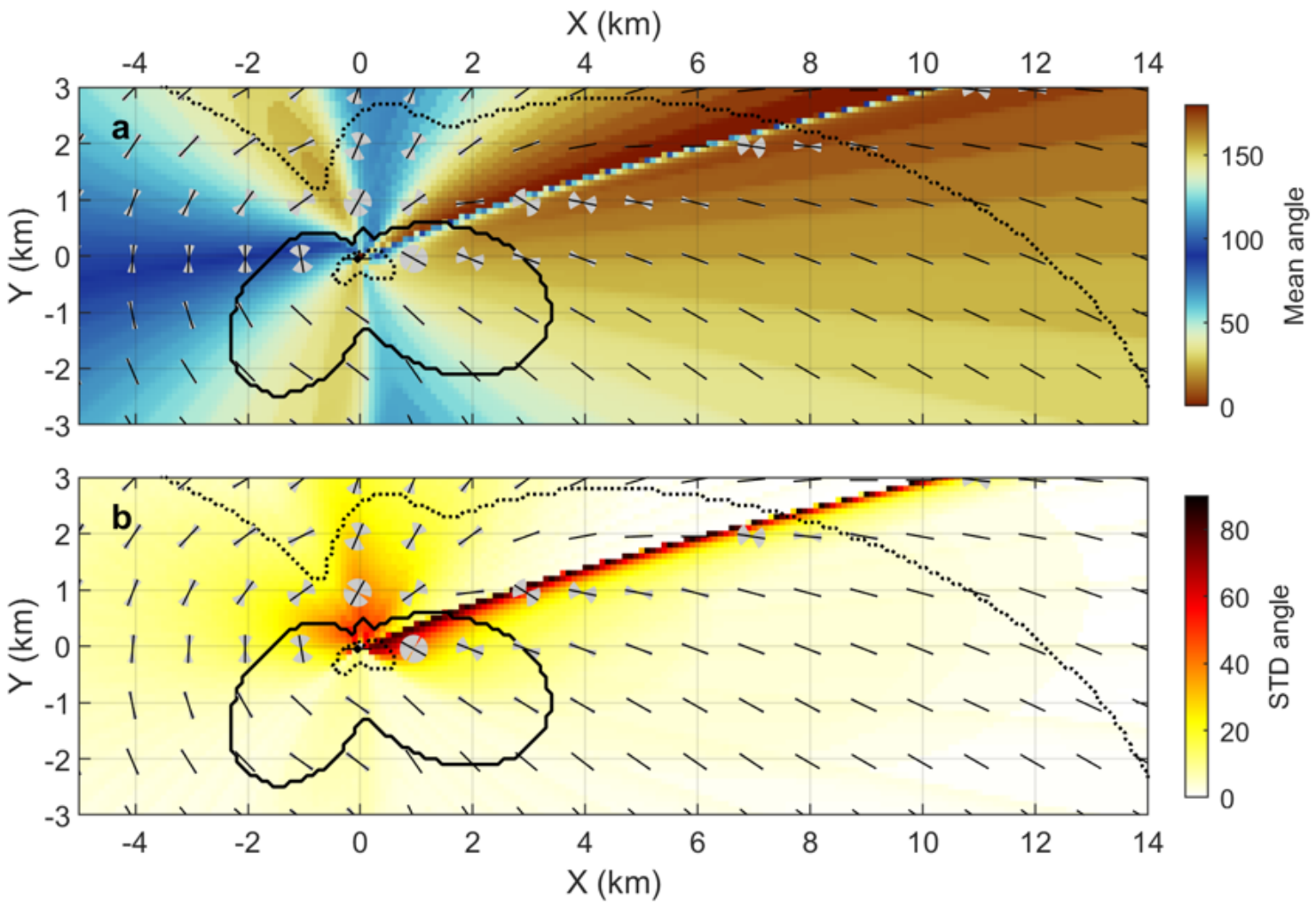

**Figure S10**: (a) Map of optimal angles for an earthquake to jump for the same setting and uncertainties as in Figure 6c. Contours in full and dotted line correspond to the probabilities of 50% and 5%, respectively. The small thin black lines and grey areas indicate the angle at which the receiver fault is optimally oriented to host a jump and their associated uncertainties. (b) Standard deviation map of the optimal angles for rupture jump. The line with high standard deviation correspond to a position where any angle of the receiver fault will produce a $V_{RSSP}$ of similar amplitude.

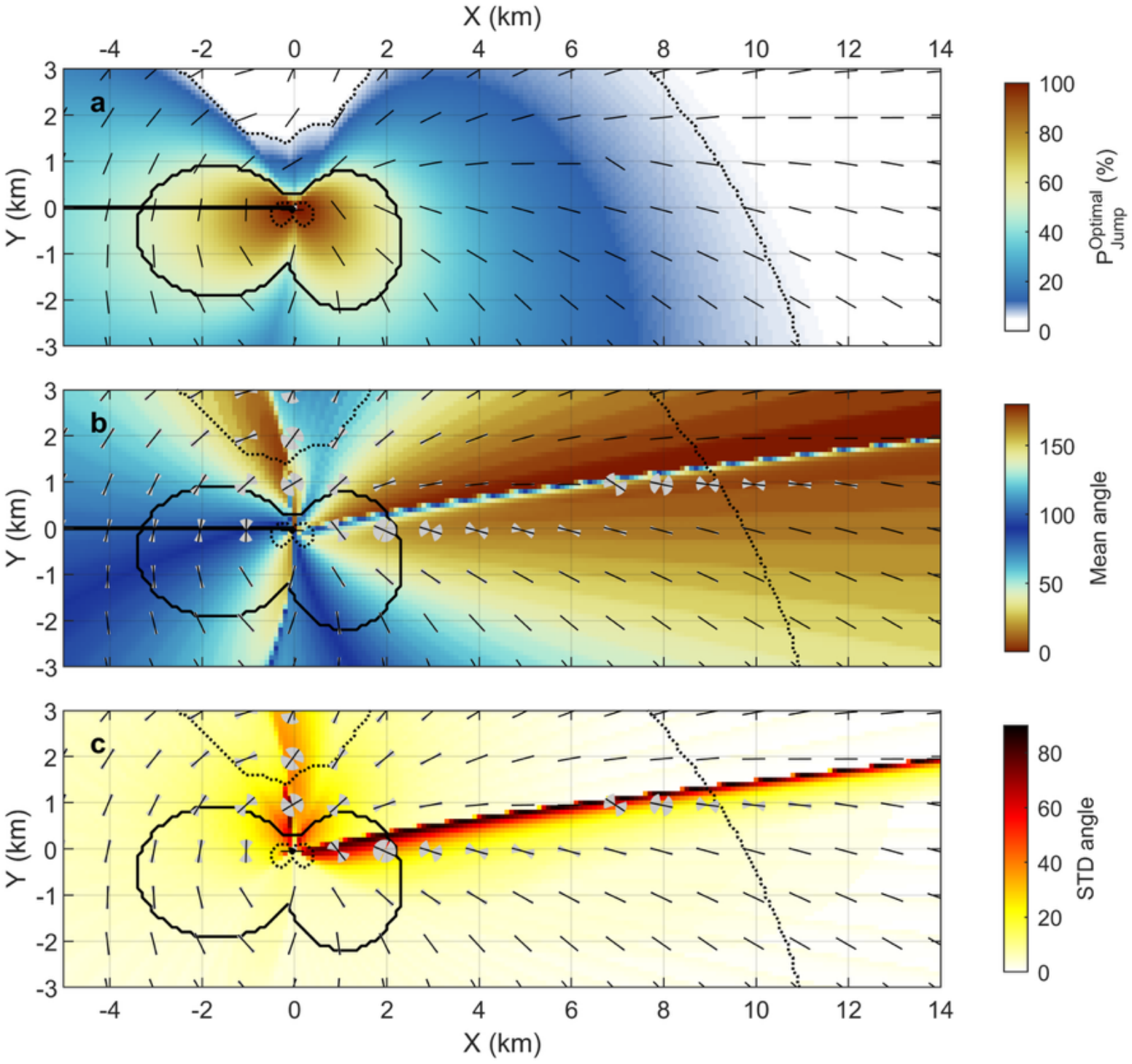

**Figure S11**: (a) Same as Figure 6.c but without the effect of orientation of the regional stress on $\sigma_0$. (b) Map of optimal angles for an earthquake to jump for the same setting and uncertainties as in Figure 6c. Contours in full and dotted line correspond to the probabilities of 50% and 5%, respectively. The small thin black lines and grey areas indicate the angle at which the receiver fault is optimally oriented to host a jump and their associated uncertainties. (b) Standard deviation map of the optimal angles for rupture jump. The line with high standard deviation correspond to a position where any angle of the receiver fault will produce a $V_{RSSP}$ of similar amplitude.

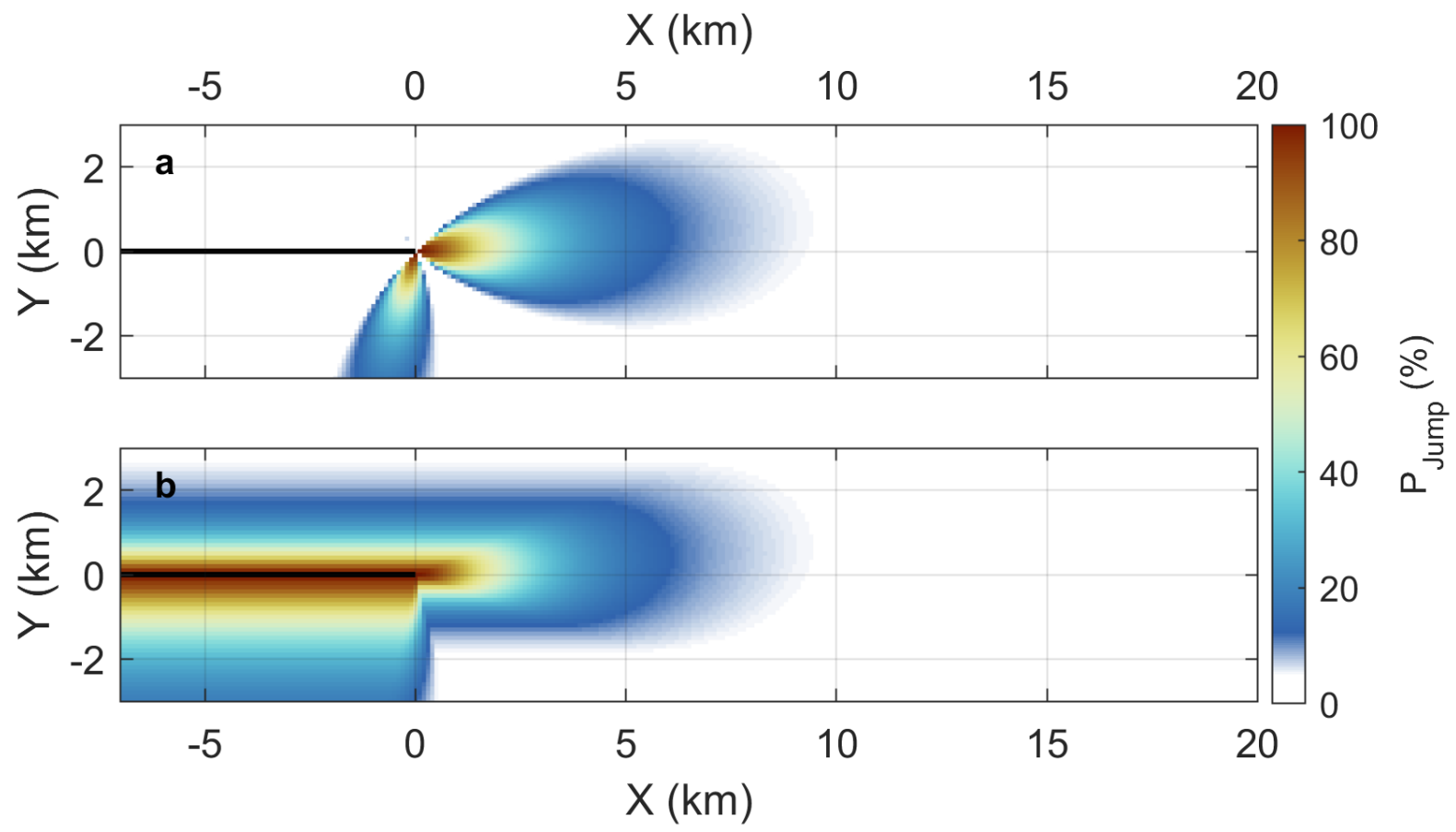

**Figure S12**: (a) Same as Figure 5.a. (b) Same as (a) but with maximum probabilities extrapolated along the fault to potentially take into account, as a maximizing scenario, the possibility that the propagating earthquake jump along the way on a receiver fault.

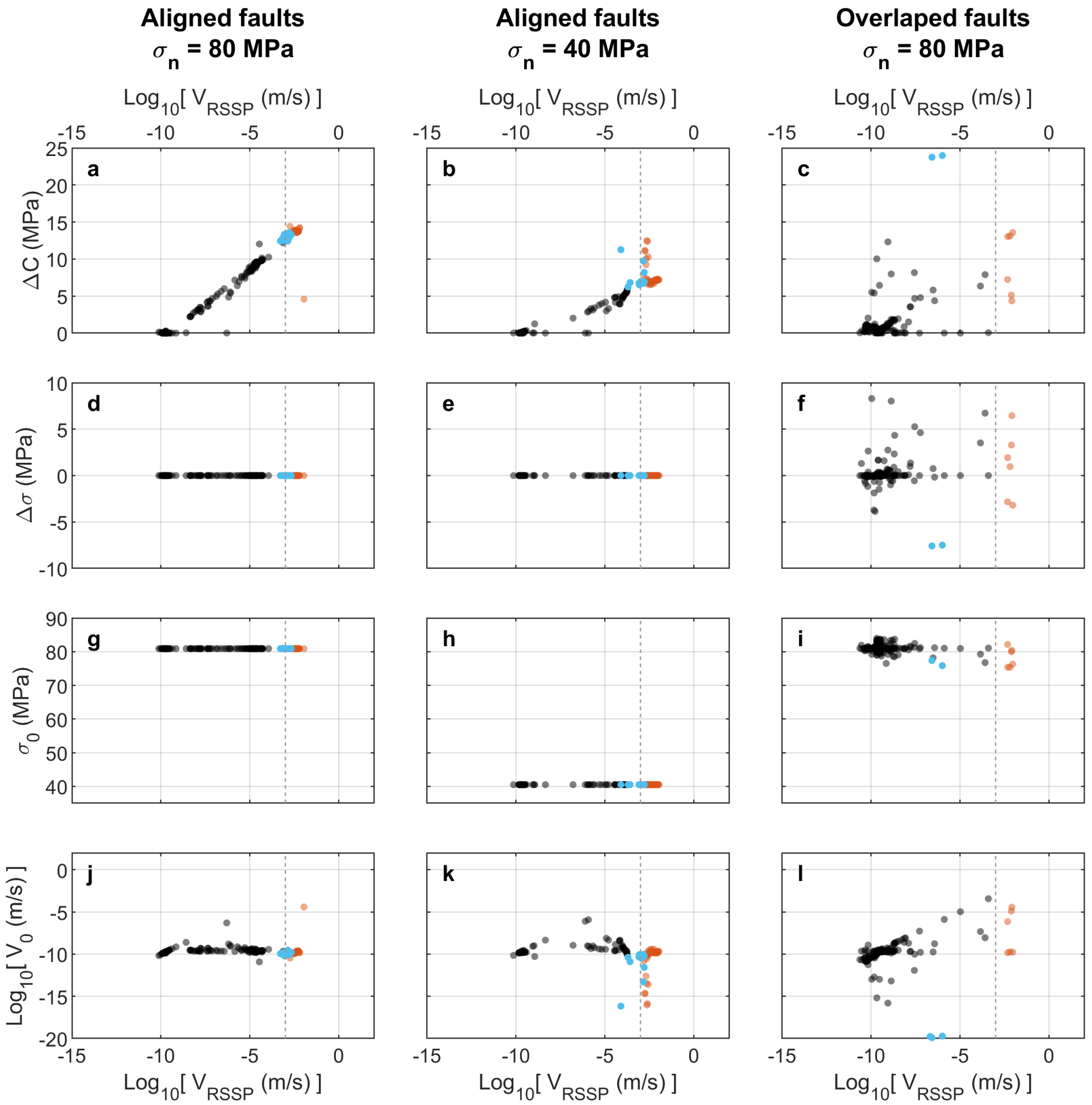

**Figure S13**: Same as Figure 3 but taking $10^{-2}$ m/s as the threshold to define an earthquake. Results from the simulations for all scenarios (Section 3). For all panels, the orange and black dots indicate events that succeeded and failed to jump, respectively. Blue dots correspond to events with high Coulomb stress change, $\Delta C$, but that did not jump. (a), (b) and (c) show the $\Delta C$ on Fault 2 due to events occurring on Fault 1 at the location of maximum $V_{RSSP}$. (d), (e) and (f) show the normal stress change, $\Delta\sigma$, on Fault 2 due to events occurring on Fault 1 at the location of maximum $V_{RSSP}$. Details on how $\Delta C$ and $\Delta\sigma$ are retrieved are in Section 3. (g), (h) and (i) show the effective normal stress, $\sigma_0$, on Fault 2 just before the start of events on Fault 1 at the location of maximum $V_{RSSP}$. (j), (k) and (l) show the slip rate, $V_0$, on Fault 2 just before the start of events on Fault 1 at the location of maximum $V_{RSSP}$.

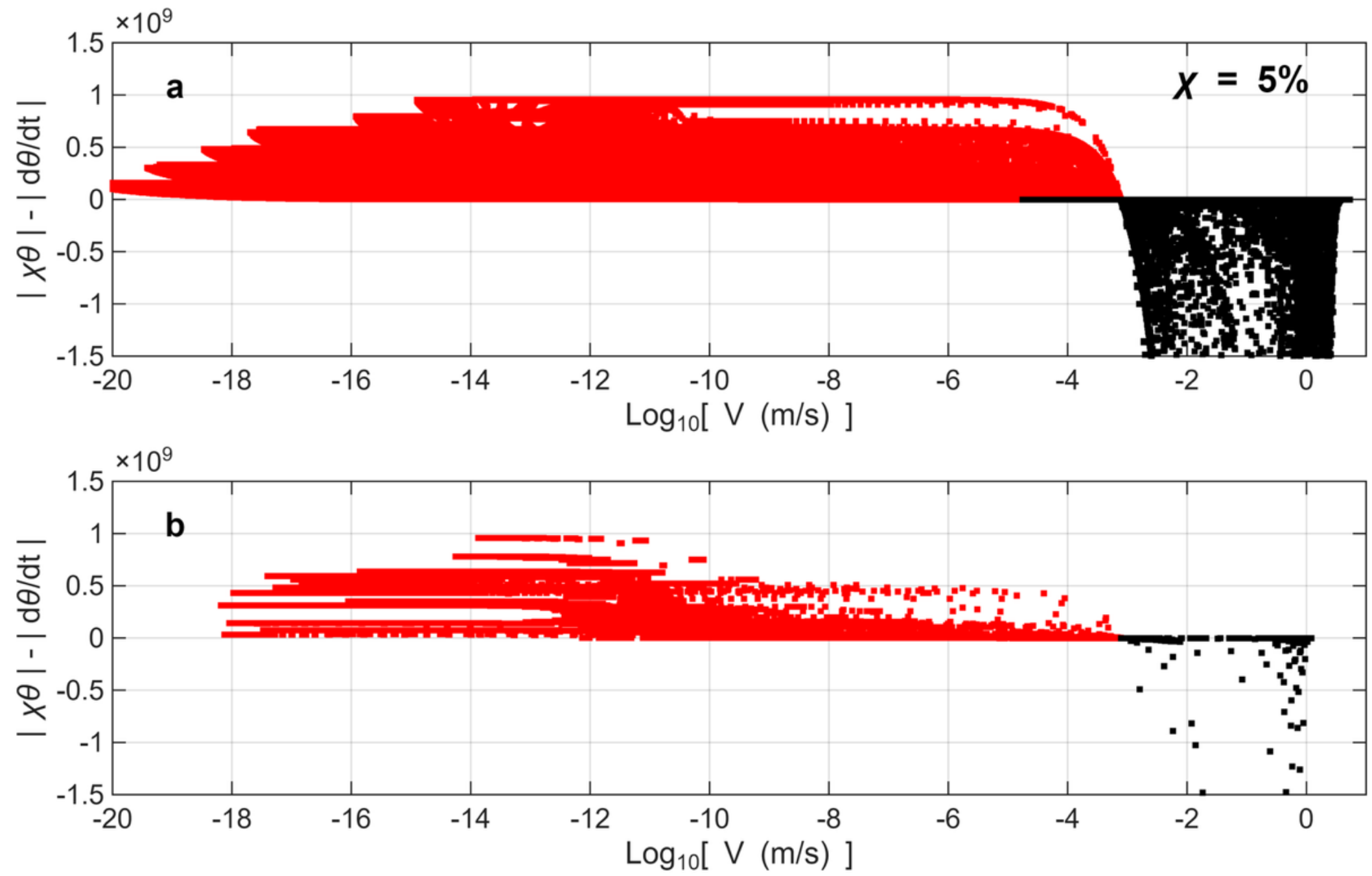

**Figure S14**: (a) Evolution of $|\chi\theta| - \left|\frac{d\theta}{dt}\right|$ as a function of slip rate $V$ in one simulation of scenario 1, for all velocity weakening subfaults on Fault 2. A change of $\left|\frac{d\theta}{dt}\right|$ is defined as *small* when it is less than a fraction $\chi$ of the state variable $\theta$. In this analysis, we set $\chi = 5\%$. *Small* changes in $\left|\frac{d\theta}{dt}\right|$, according to our definition, occur when $|\chi\theta| - \left|\frac{d\theta}{dt}\right| > 0$ (red dots), while *large* changes occur when $|\chi\theta| - \left|\frac{d\theta}{dt}\right| < 0$ (black dots). (b) Same as panel (a) but restricted to the periods near the onset of seismic events — from half a year before event initiation to a tenth of the total event duration.